\documentclass{article}

\usepackage{float}

\usepackage[normalem]{ulem}
\usepackage[utf8]{inputenc} 
\usepackage[backend=biber,style=nature, intitle = true]{biblatex}
\usepackage[verbose=true,letterpaper]{geometry}
\AtBeginDocument{
  \newgeometry{
    textheight=9in,
    textwidth=6.5in,
    top=1in,
    headheight=14pt,
    headsep=25pt,
    footskip=30pt
  }
}

\usepackage{upgreek}
\usepackage[symbol]{footmisc}

\usepackage[T1]{fontenc}    
\usepackage{hyperref}       
\usepackage{url}            
\usepackage{booktabs}       
\usepackage{amsfonts}       
\usepackage{nicefrac}       
\usepackage{microtype}      
\usepackage{xcolor}
\usepackage{textcomp}
\usepackage{authblk}
\usepackage[section]{placeins}
\usepackage{array}
\usepackage{graphicx}
\usepackage{esvect}
\usepackage{amsmath}

\DeclareUnicodeCharacter{2212}{-}

\usepackage{setspace}
\title{
Physics-based 
Modeling of Pulse and 
Relaxation of High-rate Li/CF$_{x}$-SVO batteries in 
Implantable Medical Devices}

\author[1]{\textbf{Qiaohao Liang}}
\author[2,3]{\textbf{Giacomo Galuppini}}
\author[4]{\textbf{Partha M. Gomadam}}
\author[4]{\textbf{Prabhakar A. Tamirisa}}
\author[4]{\textbf{Jeffrey A. Lemmerman}}
\author[4]{\textbf{Michael J. M. Mazack}}
\author[4]{\textbf{Melani G. Sullivan}}
\author[2]{\textbf{Richard D. Braatz}}
\author[2,5,*]{\textbf{Martin Z. Bazant}}
\affil[1]{Department of Materials Science and Engineering, Massachusetts Institute of Technology, Cambridge, MA, USA}
\affil[2]{Department of Chemical Engineering, Massachusetts Institute of Technology, Cambridge, MA, USA}
\affil[3]{University of Pavia, Pavia, PV, Italy}
\affil[4]{Medtronic Energy and Component Center, Brooklyn Center, MN, USA}

\affil[5]{Department of Mathematics, Massachusetts Institute of Technology, Cambridge, MA, USA}
\affil[*]{Corresponding author: bazant@mit.edu}

\begin{document}

\maketitle

\begin{abstract}

We present a physics-based model that accurately predicts the performance of Medtronic's implantable medical device battery lithium/carbon monofluoride (CF$_x$) - silver vanadium oxide (SVO) under both low-rate background monitoring and high-rate pulsing currents. The distinct properties of multiple active materials are reflected by parameterizing their thermodynamics, kinetics, and mass transport properties separately. Diffusion limitations of Li$^+$ in SVO are used to explain cell voltage transient behavior during pulse and post-pulse relaxation. We also introduce change in cathode electronic conductivity, Li metal anode surface morphology, and film resistance buildup to capture evolution of cell internal resistance throughout multi-year electrical tests. We share our insights on how the Li$^+$ redistribution process between active materials can restore pulse capability of the hybrid electrode, allow CF$_x$ to indirectly contribute to capacity release during pulsing, and affect the operation protocols and design principles of batteries with other hybrid electrodes. We also discuss additional complexities in porous electrode model parameterization and electrochemical characterization techniques due to parallel reactions and solid diffusion pathways across active materials. We hope our models implemented in the Hybrid Multiphase Porous Electrode Theory (Hybrid-MPET) framework can complement future experimental research and accelerate development of multi-active material electrodes with targeted performance.

\end{abstract}

\section*{Keywords}
Composite electrode; Porous electrode model; Implantable medical device; Battery;

\section*{Introduction}

Power sources of Medtronic's implantable cardioverter-defibrillators (ICDs) \cite{schmidt2001future,untereker2016power,chen2006hybrid,mond2014cardiac} require high energy density to ensure longevity and sufficient rate capability to provide high power pulses for treating abnormal heart rhythms. Such demands have led to the design of lithium/carbon monofluoride (CF$_{x}$) - silver vanadium oxide (SVO) multi-active material porous electrode batteries \cite{gan2005dual, chen2006hybrid}, which leverage the excellent energy density of CF$_x$ and power density of SVO through the use of a CF$_x$-SVO hybrid cathode. During discharge, Li$^{+}$ insertion into both octahedral and tetrahedral sites of SVO lattice is accompanied by increased structural disorder \cite{sauvage2010structural, takeuchi2001silver, onoda2001crystal, leising1994solid, west1995lithium, crespi1995characterization, crespi2001modeling} and two parallel reactions \cite{crespi1995characterization, crespi2001modeling, leising1994solid, leifer2007nuclear, cheng2011transition}: the relatively slow displacement of Ag$^{+}$ by Li$^{+}$, and the relatively fast reduction of V$^{5+}$ to V$^{4+}$ from Li$^{+}$ intercalation,
\begin{equation}\label{eq:1}
x\text{Li}^{+} + x\text{e}^{-} + \text{Ag}_{2}^{+}\text{V}_{4}^{5+}\text{O}_{11} \rightarrow \text{Li}_{x}^{+}\text{Ag}_{2−x}^{+}\text{Ag}_{x}^{0}\text{V}_{4}^{5+}\text{O}_{11} \qquad \qquad x\in[0,2] 
\end{equation}
\begin{equation}\label{eq:2}
y\text{Li}^{+} + y\text{e}^{-} + \text{Ag}_{2}^{+}\text{V}_{4}^{5+}\text{O}_{11} \rightarrow \text{Li}_{y}^{+}\text{Ag}_{2}^{+}\text{V}_{4-y}^{5+}\text{V}_{y}^{4+}\text{O}_{11} \qquad \qquad y\in[0,4] 
\end{equation}
The addition of CF$_x$ \cite{greatbatch1996lithium, davis2007simulation} significantly boosts the energy density of the multi-active material cathode, through the irreversible reaction \cite{zhang2015progress,gomadam2007modeling, tiedemann1974electrochemical, davis2007simulation}, 
\begin{equation}\label{eq:3}
x\text{Li}^{+} + x\text{e}^{-} + \text{CF}_{x}  \rightarrow x\text{LiF} + \text{C} \qquad \qquad x\in[0.8, 1.1] 
\end{equation}
The CF$_x$-SVO cathode allows the ICDs to operate smoothly under low-rate background monitoring ($\sim$$11$ $\upmu$ A, equivalent to $8$$\times$$10^{-6}$ C or $1.3$$\times$$10^{-7} \ \text{A}$$\cdot$$\text{cm}^{-2}$) for years without replacement and most critically, provide high-rate defibrillation pulses ($\sim$3.5 A, equivalent to $\sim2.5$ C or $\sim0.04 \ \text{A}$$\cdot$$\text{cm}^{-2}$) on demand. Such high-rate pulses are required to quickly charge capacitors \cite{untereker2016power, bock2012batteries}, which issues shocks to realign the patient’s heartbeat. As a result, Li/CF$_x$-SVO batteries have been widely used as power sources for today's programmable implantable medical devices that simultaneously support telemetry, monitor vital signs, and deliver 
lifesaving therapies, including ICDs, neurostimulators, and drug delivery pumps \cite{schmidt2001future, untereker2011power, chen2006hybrid}.

Researchers have attempted to develop electrochemical models of Li/CF$_x$-SVO and Li/SVO based on Porous Electrode Theory (PET) \cite{newman1975porous, newman1962theoretical, doyle1993modeling, fuller1994simulation, doyle1996comparison, smith2017multiphase} to assist in ICD battery design and to minimize the reliance on lengthy experimental tests for predicting battery performance by deploying digital twins. However, the existing models have limitations that prohibit them from accurately predicting ICD battery performance under both low-rate background monitoring and high-rate defibrillation pulsing current. Both Gomadam et al.\  \cite{gomadam2007modeling} and Strange et al.\  \cite{strange2011physics} have tried to extrapolate electrode performance from single particle dynamics, but overlooked the impact of reaction heterogeneity across SVO and CF$_x$ particle populations on battery voltage. While our previous Hybrid-MPET models \cite{liang2023hybrid} managed to capture the evolution of reaction heterogeneities across the particle population and predict voltage overshoot phenomena in Li/CF$_x$-SVO and Li/SVO batteries using many-particle models, our models and the aforementioned single particle models were evaluated only under low-rate constant current discharge operations but not under high-rate pulsing operation. As a result, the impact of solid diffusion, conductivity, and other electrode property changes on battery high-rate pulse performance have yet to be systematically investigated. To our knowledge, modeling both the high-rate defibrillation pulses and low-rate background monitoring discharge of ICD batteries has been rarely explored in the literature. Gomadam et al.\ \cite{gomadam2008modeling} attempted to capture the battery pulse and relaxation voltages during high-rate pulses through introduction of pseudo-capacitance, but the capacitance values used to fit transient voltage behavior were unrealistically large for the porous electrodes in ICDs batteries and increased model parameterization difficulty. 

\begin{figure}[!h]
  \centering
  \includegraphics[width = 12cm]{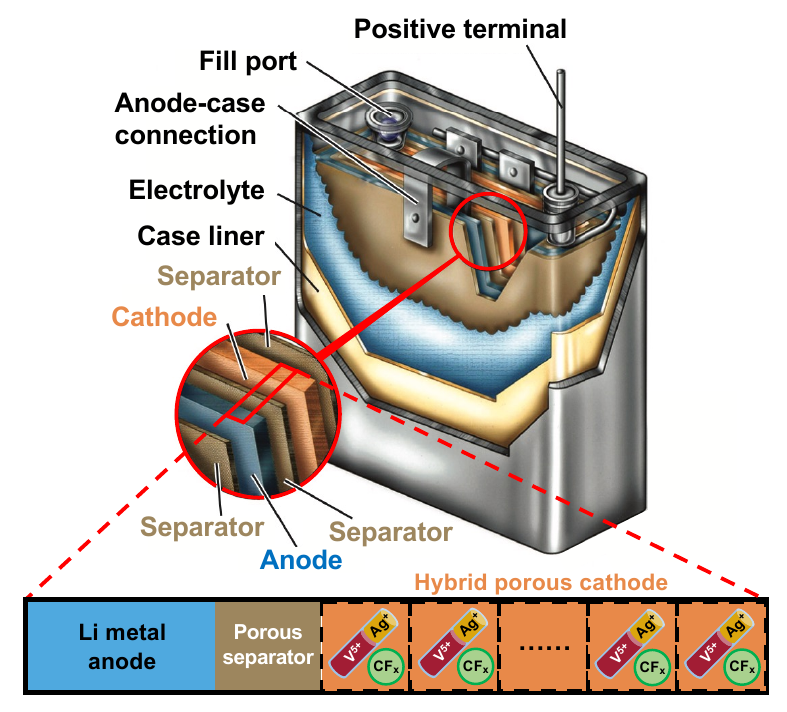}
  \vspace{-0.3cm}
  \caption{Schematic drawing of Hybrid-MPET \cite{liang2023hybrid} model formulation for Medtronic's high-rate Li/CF$_x$-SVO defibrillator battery. The cutaway view is adapted from Untereker et al.\  \cite{untereker2016power}. In our model, the electrochemical cell includes a hybrid porous cathode that contains both cylindrical SVO and spherical CF$_x$ particles, a porous separator, and a Li metal anode.}
  \label{fig:overview}
\end{figure}

In this work, we present a physics-based model of Medtronic's high-rate Li/CF$_x$-SVO ICD batteries that accurately captures their performance under both low-rate background monitoring and high-rate defibrillation pulsing. Our models are created using the open-source Multiphase Porous Electrode Theory for hybrid electrodes (Hybrid-MPET) simulation framework \cite{liang2023hybrid} (Fig.\ \ref{fig:overview}), which was previously developed to model batteries with porous electrodes that contain multiple active materials. Our models are validated against Medtronic's Li/CF$_x$-SVO electrical pulse dataset, containing 81 Li/CF$_x$-SVO cells that underwent multi-year low-rate background monitoring and quarterly high-rate defibrillation pulsing, both controlled by adjustable external loads. It is the first time such unique and considerable amount of experimental battery data is leveraged to develop and evaluate a physics-based model of Li/CF$_x$-SVO cells operating at such high current rates. To validate our models, we developed a new resistive load operation mode for the Hybrid-MPET framework, where control of cell terminal voltage and cell current is achieved through an external resistive load. Such an operation mode is more widely used in consumer electronic products compared to constant current or constant voltage, but has not been offered in other open-source mathematical battery simulation frameworks \cite{torchio2016lionsimba,sulzer2021python,berliner2021methods,albertus2007introduction, smith2017multiphase}. Instead of using large pseudo-capacitance values, we show how transient behavior of cell voltage during high-rate pulses and post-pulse relaxation can be captured by solid-phase diffusion limitations in SVO \cite{chen2006hybrid, leising1994solid, crespi1995characterization, ramasamy2006discharge,takeuchi2001silver,lee2006electrochemical,strange2011physics}.  Our model describes active material interactions in CF$_x$-SVO electrodes in the form of Li$^+$ redistribution, and successfully capture phenomena previously only observed in experimental studies \cite{chen2006hybrid}: (1) SVO providing the majority of the current during high-rate pulsing, and (2) recharge of SVO by CF$_x$ through Li$^{\text{+}}$ redistribution during post-pulse relaxation. 

To better capture evolution of Li/CF$_x$-SVO cell internal resistance over the course of the pulse test, we also introduce several property changes and degradation effects into our physics-based models: (1) increase in CF$_x$-SVO cathode solid-phase electrical conductivity \cite{chen2006hybrid, leising1994solid, crespi1995characterization, ramasamy2006discharge} due to introduction of metallic silver from Ag$^{\text{+}}$ reduction. (2) buildup of film resistance at the Li metal anode \cite{crespi1995characterization, crespi2001modeling, leising1994solid, ramasamy2006discharge} over time due to cathode dissolution \cite{bock2013kinetics, bock2013silver, li2017synthesis,demayo2015cathode, bock2015mapping,root2011resistance,li2017synthesis,le2022ag2v4o11}. (3) change in Li metal anode surface morphology \cite{hobold2023beneficial, boyle2021corrosion} under high current density discharge pulses. Using average and minimum voltage during high-rate pulsing as metrics, our model generalizes well against Medtronic's Li/CF$_x$-SVO ICD battery pulse dataset, obtaining root-mean-square error (RMSE) of 0.03 V on both pulse minimum and average voltage. We further discuss the impact of our modeling efforts on advancing understanding of the Li$^{+}$ redistribution process uniquely seen in multi-active material electrodes: Li$^{+}$ redistribution between active materials allow CF$_x$ to restore the high-rate pulsing capability of SVO and indirectly contribute to the capacity released from pulsing, which suggests that the CF$_x$-SVO electrode should be able to retain most of pulse performance with reduced amount of SVO. We also share our insights on multi-active material electrode design principles, and the challenges in model parameterization and experimental characterization of batteries with multi-active material electrodes. 

The paper is organized as follows: The Data Collection section provides more background on the specifications of Medtronic's high-rate Li/CF$_x$-SVO batteries and collection of the electrical pulse dataset. The Model section describes the formulation of the new Li/CF$_x$-SVO model in the Hybrid-MPET framework, the introduction of solid-phase diffusion in SVO, and the aforementioned electrode property change and cell degradation effects. The Results section showcases the model's ability to accurately predict pulse performance of Li/CF$_x$-SVO cells, and how the evolution of individual active material depth of discharge during both pulse and background discharge unveil their contribution to cell current at under different operations. The Discussion section expands on the role of Li$^{+}$ redistribution in the design principles of CF$_{\text{x}}$-SVO as well as general multi-active material electrodes. The Conclusion section provides a summary and an outlook for future research developments.

\section*{Data Collection}

In this work, 81 high-rate defibrillator batteries of the same proprietary design and production batch take the form of a wounded structure as seen in Fig.\  \ref{fig:overview}, where a layered assembly of a Li metal anode, a porous separator, and a CF$_x$-SVO cathode is tightly rolled, compressed, and then fit within the confines of the battery case. The CF$_x$-SVO cathode is prepared from a mixture of the active materials CF$_{\text{x}}$ and SVO (prepared by a solid-state combination reaction from $\text{Ag}_{2}\text{O}$ and $\text{V}_{2}\text{O}_{5}$ \cite{crespi1993silver, takeuchi2003advanced}), and inactive materials polytetrafluoroethylene binder and carbon black conductivity aid. All cathodes are formed by pressing a quantity of cathode mixture onto titanium current collectors. These batteries leverage a non-aqueous electrolyte LiAsF$_6$ propylene carbonate (PC)/dimethoxyethane (DME) to ensure good liquid-phase conductivity. Once the batteries are produced, they have to first go through a ``burn-in'' process at ambient temperature 37$^\circ$C that conditions the cells for optimal performance and helps with early failure detection: (1) discharge at 10 mA for 2 hr, releasing 20 mAh of cell capacity, (2) open-circuit relaxation for 11.25 hr, (3) a series of four 10 s high-rate 1.16 A pulses, with 15 s of rest between each pulse, (4) open-circuit rest for 18.5 hr and measure open-circuit voltage (OCV). The ``burn-in'' process extracts in total 32.9 mAh from each battery before pulse tests are conducted.

\begin{figure}[!h]
  \centering
  \includegraphics[width = \columnwidth]{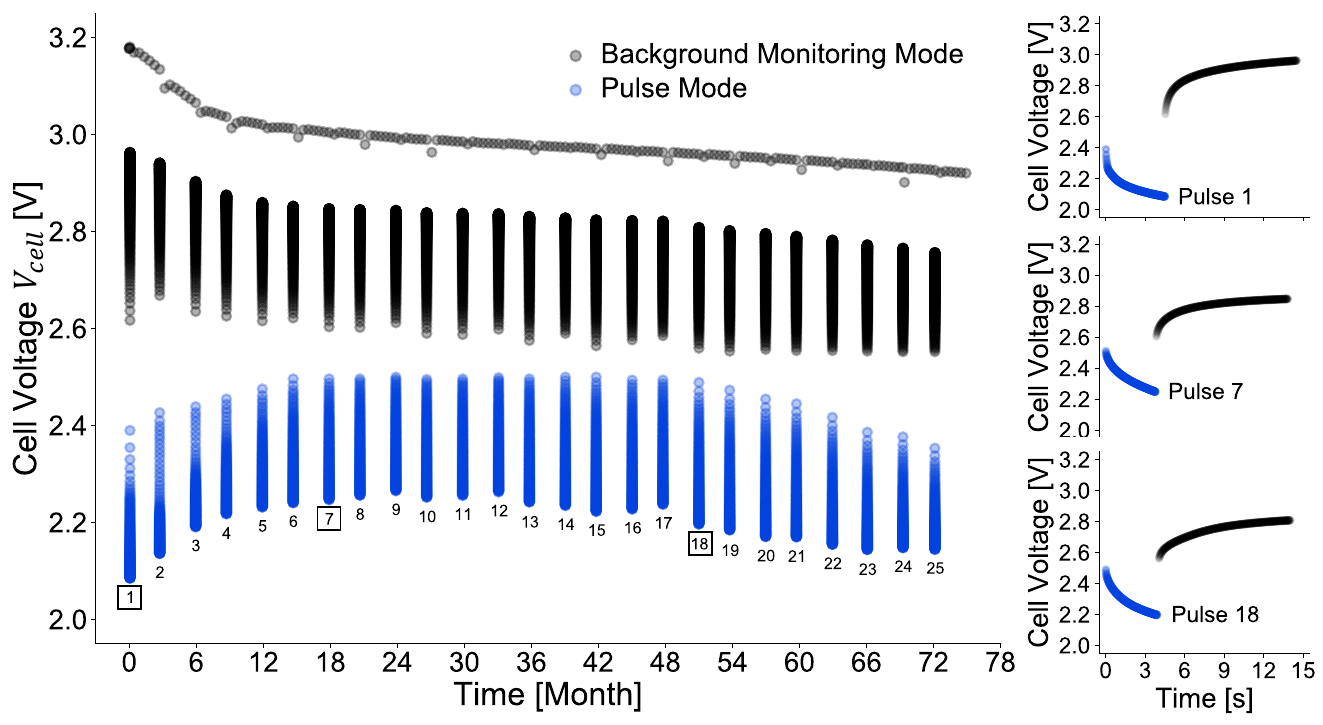}
  \vspace{-0.7cm}
  \caption{Pulse test data from a representative high-rate Li/CF$_x$-SVO defibrillator battery that ranked 50$^{\text{th}}$ percentile by average pulse minimum voltage. The black markers indicate background discharge, and the blue markers show the pulse waveforms of the battery. This representative battery reaches $\sim$56$\%$ cell depth of discharge after 78 months of quarterly 1-pulse pulse test. The pulse waveform and initial 10 s of post-pulse 
  ``relaxation'' from pulse 1, 7, and 18 are illustrated as examples, which are shown versus time that reference their respective pulse start times.}
  \label{fig:median_sample_data}
\end{figure}

If the batteries pass the ``burn-in'' test, they are placed on resistive load discharge at ambient temperature 37$^\circ$C for the pulse test. The cell terminal voltage and cell current are controlled through a resistive load $R_{\text{ext}}(t)$,
\begin{equation}\label{eq:4}
V_{\text{cell}}(t) = I_{\text{cell}}(t) \cdot R_{\text{ext}}(t)
\end{equation}
The recorded $R_{\text{ext}}(t)$ is one of the key inputs to the Hybrid-MPET model that dictates battery operation mode at any given time during testing. As seen in Fig.\  \ref{fig:median_sample_data}, the batteries undergo a quarterly 1-pulse test protocol that involves repeatedly alternating between background monitoring mode and pulse mode: (1) during low-rate background monitoring, the resistive load is kept at 270 k$\Omega$ for 3 months, yielding $\sim$$11$ $\upmu \text{A}$ background current ($8$$\times$$10^{-6}$ C or $1.3$$\times$$10^{-7} \ \text{A}$$\cdot$$\text{cm}^{-2}$). The cell voltage is measured at least once every month. (2) At the end of 3-month background monitoring mode, the resistive load is adjusted to 0.65 $\Omega$. The battery is under pulse mode ($\sim$3.5 A, equivalent to $\sim2.5$ C or $\sim0.04 \ \text{A}$$\cdot$$\text{cm}^{-2}$)) until it releases 32 J energy or average pulse voltage falls below 1 V. The average pulsing duration is $\sim$$4$ s, but varies depending on depth of discharge, pulse history, and battery-by-battery variations. Such pulsing operation resembles the typical use conditions for power sources of ICD devices when treating abnormal heartbeat on demand. The cell voltages under pulse mode and the initial 10 s of post-pulse relaxation are recorded every 0.02 s. Thus, the post-pulse ``relaxation'' is not under open circuit, but rather part of the subsequent 3 months of background monitoring mode operation controlled by 270 k$\Omega$ resistive load. At first glance, we observe that the cell pulse minimum voltage varies considerably: generally increasing from pulse 1 to pulse 9, and then decreasing from pulse 9 to pulse 25, indicating evolution of cell resistance as the pulse test proceeds. Comparing the showcased pulse 1, 7, 18, the difference in pulse minimum voltage, pulse waveform shape, and recovered voltage in initial 10 s of post-pulse relaxation also strongly infer change of electrode and material properties. Such property changes need to be accounted for by the physics-based model for better battery performance prediction across vastly different time and current scales between background monitoring and pulse mode. 

Our pulse test dataset contains information of cell voltage $V_{\text{cell}}(t)$, current $I_{\text{cell}}(t)$, resistive load $R_{\text{ext}}(t)$, and cumulative released capacity $Q_{\text{out}}(t)$ at each measured timestep for both background monitoring mode and pulse mode measurements. We have 1521 total pulses across 81 Li/CF$_x$-SVO batteries, at most covering first 6 years of battery lifetime. Due to battery-by-battery variation from manufacturing, difference in battery pulsing performance across the 81 batteries is noticeable and showcased in Supplementary Fig.\ 1. As a result, besides nominal and measured properties, the rest of our model parameterization is developed from the pulse test data of the representative battery in Fig.\  \ref{fig:median_sample_data} to capture the median battery pulsing performance across dataset. The model parameterization is also optimized with emphasis towards prediction accuracy during high-rate pulse and post-pulse relaxation, with specific interest in the following metrics that describe Li/CF$_x$-SVO battery pulse performance: pulse minimum voltage $V^{\text{min}}_{\text{pulse}}$ and pulse average voltage $V^{\text{avg}}_{\text{pulse}}$, as well as the average voltage $V^{\text{avg}}_{\text{relax}}$ and end voltage $V^{\text{end}}_{\text{relax}}$ from initial 10 s of post-pulse relaxation. The $V^{\text{min}}_{\text{pulse}}$ and $V^{\text{avg}}_{\text{pulse}}$ are correlated with the power capability of the cell under high-rate pulses while $V^{\text{avg}}_{\text{relax}}$,$V^{\text{end}}_{\text{relax}}$ infer how fast the cell can recover from high-rate pulses. We use root mean square error $\mathrm{RMSE}=\sqrt{\frac{1}{P}\sum_{p=1}^{P}(V_p-\hat{V}_p)^2}$ to quantify model goodness-of-fit, where $P$ is the number of pulses, $\hat{V}_p$ and $V_p$ are the model predictions and experimental measurement at the $p^{\text{th}}$ pulse, respectively. The model is then evaluated on the rest of the 80 cells in the dataset, and its accuracy quantified on the same metrics.

\section*{Model}

Compared to our previous Hybrid-MPET models \cite{liang2023hybrid} of medium-rate Li/SVO and Li/CF$_x$-SVO batteries, our Hybrid-MPET model for high-rate Li/CF$_x$-SVO batteries keeps the intra-particle and inter-particle scale parallel reactions from the active Ag$^{+}$, V$^{5+}$, and CF$_x$, but no longer neglects the solid-phase diffusion limitations. We also model several property changes and degradation effects documented in literature to better capture battery pulse performance, including changes in cathode conductivity, anode surface morphology after repeated high-rate pulsing, and anode film resistance build-up from cathode dissolution. 

\subsection*{Basic properties and formulation}

The basic electrode, material, and particle properties used in our model can be seen in Table \ref{table:SVOCFxelectrode}. Our model is run in an isothermal setting at ambient temperature 37$^\circ$C. The SVO material is assumed to be able to take in 6 stoichiometry of Li$^{+}$, 2 from the Ag$^{+}$ reaction and 4 from the V$^{5+}$ reaction, resulting in volumetric energy densities $q_{_{\text{Ag}}} = \frac{1}{3}q_{_{\text{SVO}}}$, $q_{_{\text{V}}} = \frac{2}{3}q_{_{\text{SVO}}}$. Since inserted Li$^{+}$ coexist in SVO, the particle volume and active reaction area is assumed to be shared across the Ag$^{+}$ and V$^{5+}$ reactions, and thus the capacity fraction weighted SVO filling fraction \cite{liang2023hybrid} is $\overline{c}_{_{\text{SVO}}} = \frac{1}{3}\overline{c}_{_{\text{Ag}}} + \frac{2}{3}\overline{c}_{_{\text{V}}}$. It is  estimated that $x\approx 1$ in CF$_x$ \cite{gomadam2007modeling}, and CF$_x$ particles are assumed to consist of a single CF phase. The Li/CF$_x$-SVO batteries in this study have the same CF$_{\text{x}}$-SVO cathode material, where the nominal capacity ratio between CF$_x$ and SVO is 2.031:1, yielding nominal capacity fractions $\frac{1}{3}$ and $\frac{2}{3}$ for SVO and CF$_x$, and a nominal specific energy density of 434.12 mAh$\cdot \text{g}^{-1}$ for the electrode. For the representative battery, the measured electrode thickness is 140 $\upmu$m and measured electrode mass 3.055 g. Other batteries in the dataset have slight difference in cathode thickness, mass, and estimated active solid volume fraction and capacity due to manufacturing variation.

\renewcommand{\thetable}{1}
\renewcommand{\arraystretch}{1.2}
\newcolumntype{P}[1]{>{\centering\arraybackslash}p{#1}}
\begin{table}[!h]
  \caption{Electrode and materials properties of CF$_{\text{x}}$-SVO multi-active material porous cathode}
  \centering
  \begin{tabular}{ccccccc}
  \hline
   Parameter &Description & Unit & \multicolumn{4}{c}{CF$_x$-SVO Electrode}
   \\
  \hline
    $L$ &  measured electrode thickness & $\upmu$m &    \hspace{+55pt} 140
    \\
    $A$ & nominal electrode area & cm$^{-2}$ &    \hspace{+55pt} 84.5 
    \\
    $m$ & measured electrode mass & g &    \hspace{+55pt} 3.0549 
    \\
    $\epsilon$ & nominal electrode porosity & -- & \hspace{+55pt} 0.33    
    \\
    $\widetilde{m}$ & nominal active solid mass fraction & -- & \hspace{+55pt} 0.9
    \\
    $P_{L}$ & estimated active solid volume fraction & -- & \hspace{+55pt} 0.95
    \\
    $\tau$ & tortuosity & -- & \hspace{+55pt} $\epsilon^{-0.85}$
    \\
    $\widetilde{Q}_{\text{SVO}} $ & nominal capacity fraction of SVO & --  & \hspace{+55pt} 0.333
    \\
    $\widetilde{Q}_{\text{CF}_x}$ & nominal capacity fraction of $\text{CF}_x$ & --  & \hspace{+55pt} 0.667
    \\ 
    $\rho_{\text{e}, c}$ & nominal specific energy density & mAh$\cdot$$\text{g}^{-1}$  & \hspace{+55pt} 434.12
    \\
    $Q_{c}$ & estimated capacity & mAh  & \hspace{+55pt} 1326
    \\

  \hline
        Parameter &Description & Unit & \multicolumn{3}{c}{\hspace{+0pt} SVO}  & CF \\
  \hline
    $\rho$ & density & g$\cdot$cm$^{-3}$  & \multicolumn{3}{c}{\hspace{+0pt} 4.789}  & 2.75
    \\
    $\rho_{\text{e}}$ & specific energy density & mAh$\cdot$g$^{-1}$ & \multicolumn{3}{c}{\hspace{+0pt} 270} & 865 
    \\
    $q$ & volumetric energy density & mol$\cdot$m$^{-3}$ & \multicolumn{3}{c}{\hspace{+0pt} 48322} & 88739 
    \\
    $R$ & particle radius & $ \upmu$m  & \multicolumn{3}{c}{\hspace{+0pt} $\mathcal{N}$(0.835, 0.04)}  & $\mathcal{N}$(1, 0.05)
    \\
    $h$ & particle length & $ \upmu$m  & \multicolumn{3}{c}{\hspace{+0pt} 16.7}  & --
    \\

    
    \hline
    \end{tabular}
  \label{table:SVOCFxelectrode}
\end{table}
For electrolyte transport, a concentrated Stefan-Maxwell electrolyte model \cite{newman2021electrochemical, smith2017multiphase} is used to represent the high-rate electrolyte 1.1M LiAsF$_6$ PC/DME 1:1. The properties of the binary electrolyte, including liquid-phase diffusivity $D_{l}(c_{_{l}}, T)$, ionic conductivity $\sigma_{l}(c_{_{l}}, T)$, transference number $t^{0}_{+}(c_{_{l}}, T)$, and thermal factor $\gamma(c_{_{l}}, T)$, are obtained from a corresponding Advanced Electrolyte Model (AEM) \cite{gering2006prediction,gering2017prediction, logan2018critical}, and the exact values are shown in Supplementary Fig.\ 2.

The basic model formulation follows the schematic drawing in Fig.\  \ref{fig:overview}. The anode is a Li-foil (Li metal) anode foil and we assume the cell is cathode capacity limited. The porous separator used has thickness 20 $\upmu$m, porosity 0.4, tortuosity $\epsilon^{-0.85}$, and is discretized into 5 finite volumes. The CF$_x$-SVO cathode is spatially discretized into $N$ = 5 finite volumes, each containing $P_{_{\text{SVO}}}=5$ cylindrical SVO and $P_{_{\text{CF}}}=5$ spherical CF$_x$ particles with sizes sampled from their respective distributions. 
To simulate a multi-active material electrode with limited number of particles and still ensure correct capacity fractions $\widetilde{Q}_{\text{CF}_{x}}$, we need to preserve $\widetilde{Q}_{\text{SVO}}$ at both electrode scale and finite volume scale, and thus introduce volume correction terms \cite{liang2023hybrid} to each finite volume $\mathcal{V}_{n}$ such that volume fractions $\widetilde{V}_{n, p_{_{\text{CF}}}}$, $\widetilde{V}_{n, p_{_{\text{SVO}}}}$ reflect correct active material contributions to the average state of charge and total reaction rate when the multi-active material electrode CF$_{\text{x}}$-SVO has capacity fractions $\widetilde{Q}_{\text{CF}_{x}}$, $\widetilde{Q}_{\text{SVO}}$. As for the cell depth of discharge, i.e.\ cathode filling fraction, it can be both predicted from the model by averaging across finite volumes $\overline{c} = \frac{1}{N} \sum_{n=1}^{N}\overline{c}_{n}$ and estimated at each measured timestep from experimental data using $\bar{c}(t) = \frac{Q_{out}}{Q_{c}} = \frac{Q_{out}}{\rho_{\text{e}, c}m}$. 
The initial depth of discharge of the active materials are set to $c_{_{\text{Ag}}}(t=0) = 0.0383$, $c_{_{\text{V}}}(t=0) = 0.0356$, and $c_{_{\text{CF}}}(t=0) = 0.0156$ across the particles to account for cell capacity release from burn-in process ($\overline{c}(t=0) = 0.024$) and ensure initial potential balance between active materials at 3.18 V, which matches the cell voltage first measured at the start of the pulse test for the representative battery.

\subsection*{Reaction kinetics}

With only minor adjustments to the exchange-current density rate constants, we adopt the Butler-Volmer (BV) \cite{newman2021electrochemical,bard2001fundamentals,dreyer2016new,heubner2015investigation, bazant2013theory, thomas2002mathematical} reaction kinetics of Ag$^{+}$, V$^{5+}$ separately from our previous models of Li/SVO and Li/CF$_x$-SVO \cite{liang2023hybrid} to obtain their reaction current densities $i$ as functions of both electrolyte concentration and solid-phase filling fractions,
\begin{equation}\label{eq:5}
i_{_{\text{Ag}}} = j_{_{\text{Ag}}} e = k_{_{\text{Ag}}} (c_{_{l}})^{0.5} c_{_{\text{Ag}}}^{0.1}(1-c_{_{\text{Ag}}})^{5.5}\!
\left[\exp\!\left(-\frac{0.5 e \eta_{_{\text{Ag}}}}{k_{\text{B}}T}\right)\! - \exp\!\left(\frac{0.5 e \eta_{_{\text{Ag}}}}{k_{\text{B}}T}\right) \right]
\end{equation}
\begin{equation}\label{eq:6}
i_{_{\text{V}}} = j_{_{\text{V}}} e = k_{_{\text{V}}} (c_{_{l}})^{0.5} c_{_{\text{V}}}^{0.5}(1-c_{_{\text{V}}})^{0.5}\!
\left[\exp\!\left(-\frac{0.5 e \eta_{_{\text{V}}}}{k_{\text{B}}T}\right)\! - \exp\!\left(\frac{0.5 e \eta_{_{\text{V}}}}{k_{\text{B}}T}\right) \right]
\end{equation}
where $k_{_{\text{Ag}}} = 1.5$$\times$$10^{-5} \ \text{A}$$\cdot$$\text{m}^{-2}$ and $k_{_{\text{V}}} = 7$$\times$$10^{-1} \ \text{A}$$\cdot$$\text{m}^{-2}$. Note only the sides of the cylindrical SVO particle are active for reaction, and the SVO particle is further discretized in the radial direction by $\Delta r = 0.02 \ \upmu \text{m}$ to capture concentration gradients. From previous modeling studies of Li/CF$_x$-SVO batteries under low- and medium- current rates \cite{gomadam2007modeling, liang2023hybrid}, it was clear that the active material depletion order was Ag$^{+}$, V$^{5+}$, CF$_x$. While the pulse test data has quarterly pulses mixed in, we expect the thermodynamics to dominate again soon after the operation has been switched to background monitoring mode. As a result, the relatively fast decrease in cell voltage from $\sim$3.2 V to $\sim$3.0 V during the first few segments background monitoring mode in Fig.\  \ref{fig:median_sample_data} can be directly attributed to increased Ag$^{+}$ depletion, allowing us to refine our selection for its exchange current density rate constant $k_{_{\text{Ag}}}$. Compared to the conventional Li$^{+}$ intercalation reaction of V$^{5+}$, it is commonly believed that the Ag$^{+}$ displacement reaction is kinetically slower \cite{sauvage2010structural}; such difference is in turn reflected in a much smaller $k_{_{\text{Ag}}}$. The transport of metallic silver atoms as well as nucleation and growth of silver nanoparticles \cite{ramasamy2006discharge, crespi2001modeling}, which require statistical models such as Kolmogorov–Johnson–Mehl–Avrami (KJMA) theory \cite{allen2008analysis, allen2007kinetic, oyama2012kinetics}, is beyond the scope of our electrochemical porous electrode models. We thus assume they do not interfere with electrochemical reactions of Ag$^{+}$ and V$^{5+}$ since remarkably fast Ag atomic diffusion were measured during lithiation of silver doped vanadium oxide systems \cite{galy2008new} with similar layered structures. 

The electrochemical reduction reaction of CF$_x$ has been experimentally observed to be irreversible \cite{zhang2015progress} and obey Tafel kinetics \cite{tiedemann1974electrochemical, davis2007simulation} even at very low current rates, indicating a very low exchange current density rate constant for CF$_x$ \cite{hunger1975rate}. The rest of background monitoring mode segments in Fig.\  \ref{fig:median_sample_data} can be directly attributed to increased CF$_x$ depletion, and allows us to obtain a matching $k_{_{\text{CF}}}$, 
\begin{equation}\label{eq:7}
i_{_{\text{CF}}} = j_{_{\text{CF}}}f^{\text{CF}}_{A} e = k_{_{\text{CF}}}f^{\text{CF}}_{A}
c_{_{\text{CF}}}(1-c_{_{\text{CF}}})^{\frac{1}{3}}
\exp\!\left(-\frac{0.57 e \eta_{_{\text{CF}}}}{k_{\text{B}}T}\right) 
\end{equation}
where $k_{_{\text{CF}}} = 1.5$$\times$$10^{-5} \ \text{A}$$\cdot$$\text{m}^{-2}$. The small $k_{_{\text{CF}}}$ reflects the kinetics limitation of CF$_x$ \cite{tiedemann1974electrochemical, davis2007simulation, hunger1975rate} as an electrode material, which has been traditionally only used for medium-rate batteries \cite{schmidt2001future,untereker2016power,gan2005dual,chen2006hybrid,mond2014cardiac}. Its active material is assumed to be coated at the spherical CF$_{x}$ particle surface, and thus CF$_x$ has negligible mass transport limitations. The spherical $c_{_{\text{CF}}}$ can be treated as a homogeneous particle, and its active reaction area available decreases with increasing $c_{_{\text{CF}}}$ \cite{gomadam2007modeling}. Since current is the product of current density and active reaction area, the effect of an decreasing active reaction area can be introduced to $k_{_{\text{CF}}}$ as an area term $f^{\text{CF}}_{A}(c_{_{\text{CF}}}) = (1-c_{_{\text{CF}}})^{\frac{2}{3}}$ in Eq.\  \ref{eq:7}. In our model of Li/CF$_{\text{x}}$-SVO, the contributions to current from Ag$^{+}$ and CF$_{\text{x}}$ reactions during high-rate pulsing are small when compared to that of the V$^{5+}$ reaction; thus, SVO, and more specifically its fast V$^{5+}$ reduction reaction, is the source of high-power capability of Li/CF$_x$-SVO cells.

Due to coexistence of 3 reduction reactions from 2 different particles, equations for both intra- and inter-particle scale parallel reactions are used in the Hybrid-MPET framework \cite{liang2023hybrid}. The sum of reactions from all particles in $n^{\text{th}}$ finite volume $\mathcal{V}_{n}$ yields its total finite volume average reaction rate,
\begin{equation}\label{eq:8}
R^{V, \text{corr}}_{n} = -(1-\epsilon)P_{L}\!\left[
\sum_{p_{_{\text{SVO}}} = 1}^{P_{_{\text{SVO}}}} \widetilde{V}_{n, p_{_{\text{SVO}}}} \left(\frac{1}{3}\frac{\partial \overline{c}_{n,p_{_{\text{SVO}}}, _{\text{Ag}}}}{\partial t} + \frac{2}{3} \frac{\partial \overline{c}_{n,p_{_{\text{SVO}}}, _{\text{V}}}}{\partial t} \right)
+
\sum_{p_{_{\text{CF}}} = 1}^{P_{_{\text{CF}}}} \widetilde{V}_{n, p_{_{\text{CF}}}} \frac{ \partial \overline{c}_{n,p_{_{\text{CF}}}}}{\partial t}
\right] 
\end{equation} 
where the corrected volume fractions $\widetilde{V}_{n, p_{_{\text{SVO}}}}$, $\widetilde{V}_{n, p_{_{\text{CF}}}}$ \cite{liang2023hybrid} ensure that the correct capacity ratio is preserved in all finite volumes and thus also the multi-active material electrode regardless of the number of particles used \cite{liang2023hybrid}. For SVO, the particle-scale average volumetric reaction rate $\frac{\partial \overline{c}_{n,p_{_{\text{SVO}}}, _{\text{Ag}}}}{\partial t}$, $\frac{\partial \overline{c}_{n,p_{_{\text{SVO}}}, _{\text{V}}}}{\partial t}$ are derived from the volume average of reaction rate of Ag$^{+}$, V$^{5+}$ reduction from the set of cylindrical SVO shells at different particle radii \cite{smith2017intercalation, zeng2013efficient, zeng2014phase}. The macroscopic discharge current density $i_\text{cell}$ is thus supported by all 3 parallel reduction reactions at the cathode,
\begin{equation}\label{eq:9}
i_\text{cell} = -\sum_{n = 1}^{N = 5}eR^{V,\text{corr}}_{n}\frac{L}{N} 
\end{equation}

\subsection*{Thermodynamics}
\begin{figure}[!h]
  \centering
  \includegraphics[width = \columnwidth]{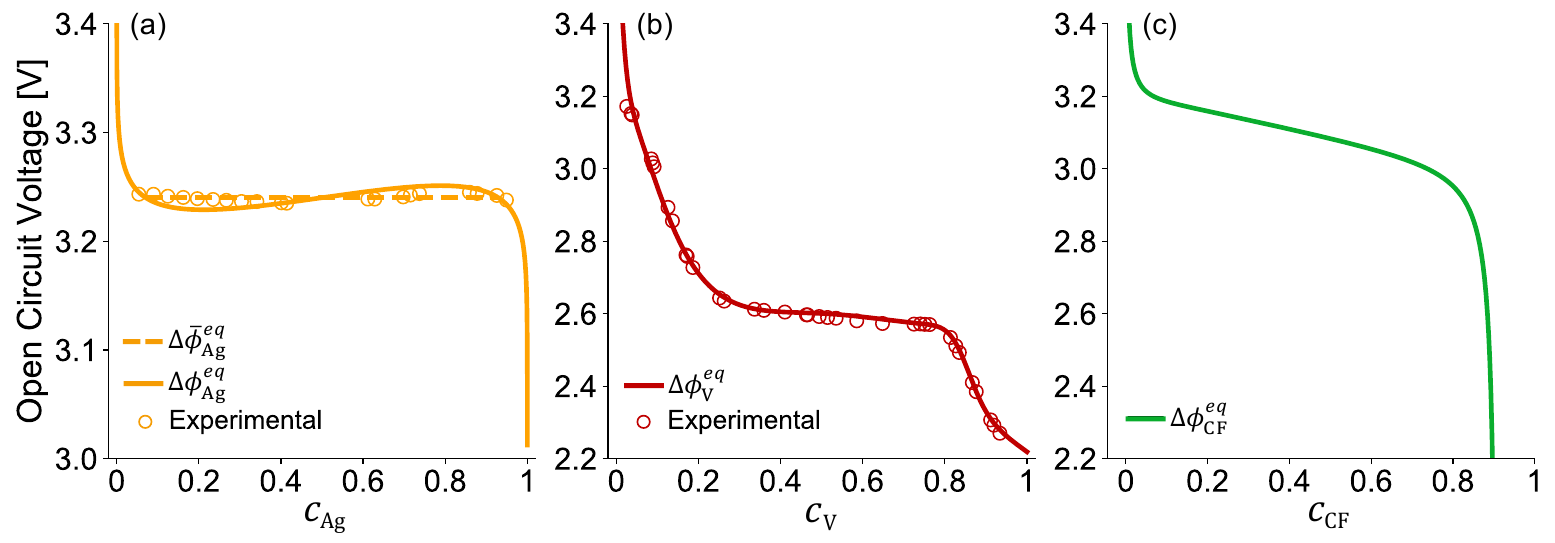}
  \vspace{-0.7cm}
  \caption{Open-circuit voltages as functions of depth of discharge for silver, vanadium, and carbon monofluoride reduction reaction. The experimentally measured OCV of SVO \cite{crespi1995characterization, crespi2001modeling, gomadam2007modeling} is split by assuming that silver and vanadium reduction are dominant in the first $\frac{1}{3}$ and last $\frac{2}{3}$ capacity of SVO, respectively. (a) The non-monotonic $\mu^{h}_{\text{Ag}}$ is derived from a double-well like free energy function, which is formulated such that a common tangent construction between its two local energy minima has a composition width that matches the width of experimentally measured voltage plateau at 3.24 V during Ag$^{+}$ reduction in the first $\frac{1}{3}$ capacity of SVO \cite{crespi1995characterization, crespi2001modeling, gomadam2007modeling}. (b) Li$^{+}$ insertion due to V$^{5+}$ reduction is modeled as intercalation into solid solution material and $\Delta\phi^{eq}_{\text{V}}$ is fitted as a function of vanadium utilization from the last $\frac{2}{3}$ capacity of the experimentally measured OCV of SVO. (c) $\Delta\phi^{eq}_{\text{CF}}$ is extracted from experimental data \cite{gomadam2007modeling} by assuming Tafel kinetics.}
  \label{fig:blackwell_OCVs}
\end{figure}

The OCV of Li/SVO \cite{crespi1995characterization, crespi2001modeling, gomadam2007modeling} has a voltage plateau at 3.24 V covering the first $\frac{1}{3}$ cell capacity, strongly inferring coexistence of multiple stable phases throughout the Ag$^{+}$ displacement reaction. As a result, to describe the thermodynamics of Ag$^{+}$ reduction as a function of its depth of discharge $c_{_{\text{Ag}}}$, we utilize a double-well \cite{cahn1961spinodal} like homogeneous free energy functional and a simple gradient penalty
term to describe the non-homogeneous free energy \cite{rowlinson1979translation, cahn1958free}. The resulting diffusional chemical potential $\mu_{_{\text{Ag}}}$ of Li$^{+}$ inserted into SVO due to Ag$^{+}$ reduction is
\begin{equation}\label{eq:10}
\mu_{_{\text{Ag}}}(c_{_{\text{Ag}}}) = \frac{\updelta G}{\updelta c_{_{\text{Ag}}}} 
= \frac{\partial g}{\partial c_{_{\text{Ag}}}} - \nabla \cdot \frac{\partial g}{\partial \nabla c_{_{\text{Ag}}}}= \mu^{h}_{_{\text{Ag}}}(c_{_{\text{Ag}}}) - \frac{\kappa}{q_{_{\text{Ag}}}} \nabla^{2} c_{_{\text{Ag}}}
\end{equation}
where
\begin{equation}\label{eq:11}
\mu^{h}_{_{\text{Ag}}}(c_{_{\text{Ag}}}) = k_{\text{B}}T \ln\!\left(\frac{c_{_{\text{Ag}}}}{1-c_{_{\text{Ag}}}}\right)\! + \widetilde{\Omega} k_{\text{B}}T (1-2c_{_{\text{Ag}}}),
\end{equation}
$k_{\text{B}} =  1.38$$\times$$10^{-23}\  \text{J}$$\cdot$$\text{K}^{-1}$, $T = 310.15\ \text{K}$, $\widetilde{\Omega}_{a}=3.0$, and $\kappa=5$$\times$$10^{-7} \ \text{J}$$\cdot$$ \text{m}^{-1}$. As seen in Fig.\  \ref{fig:blackwell_OCVs}(a), the corresponding OCV is thus 
\begin{equation}\label{eq:12}
\Delta\phi^{\text{eq}}_{\text{Ag}}(c_{_{\text{Ag}}}) = 3.24 - \frac{\mu_{_{\text{Ag}}}(c_{_{\text{Ag}}})}{e} 
\end{equation}

The OCV of Li/SVO \cite{crespi1995characterization, crespi2001modeling, gomadam2007modeling} also has a second voltage plateau at 2.6 V that spans across the rest $\frac{2}{3}$ cell capacity; most studies attribute it to the V$^{5+}$ reduction process that is accompanied by Li$^{+}$ intercalation into an single-phase host structure with high degree of disorder \cite{sauvage2010structural, crespi1995characterization, crespi2001modeling, west1995lithium}. As seen in Fig.\  \ref{fig:blackwell_OCVs}(b), the thermodynamics of V$^{5+}$ reduction is described by a separate OCV $\Delta\phi^{\text{eq}}_{\text{V}}$ as a function of its filling fraction $c_{_{\text{V}}}$,
\begin{equation}\label{eq:13}
\Delta\phi^{\text{eq}}_{\text{V}}(c_{_{\text{V}}}) = 0.823 + \exp(-80c_{_{\text{V}}}) + 
\frac{3.177 
+92.839c_{_{\text{V}}}^2
+49.148c_{_{\text{V}}}^4
-658.841c_{_{\text{V}}}^6
+589.917c_{_{\text{V}}}^8
}
{1 
+39.404c_{_{\text{V}}}^2
-6.299c_{_{\text{V}}}^4
-171.554c_{_{\text{V}}}^6
+106.016c_{_{\text{V}}}^8
+65.794c_{_{\text{V}}}^{10}}
\end{equation}

By assuming Tafel kinetics \cite{tiedemann1974electrochemical, davis2007simulation} in the form of Eq.\  \ref{eq:7}, we can extract \cite{gomadam2007modeling} its OCV $\Delta\phi^{\text{eq}}_{\text{CF}}$ from experimental data as a function of its depth of discharge $c_{_{\text{CF}}}$,
\begin{multline}\label{eq:14}
\Delta\phi^{\text{eq}}_{\text{CF}}(c_{_{\text{CF}}}) = 0.505 
-0.047\ln\!\left(c_{_{\text{CF}}}(1-c_{_{\text{CF}}})^{\frac{1}{3}}\right) \\
+\frac{
3.426
+1317.958c_{_{\text{CF}}}
+9596.395c_{_{\text{CF}}}^2
+12208.742c_{_{\text{CF}}}^3
}{
1
+527.395c_{_{\text{CF}}}
+3570.789c_{_{\text{CF}}}^2
-4231.792c_{_{\text{CF}}}^3
-361.622c_{_{\text{CF}}}^4
} 
\end{multline}

\subsection*{Solid diffusion in SVO}
From our modeling perspective, the cell voltage is a macroscopic and collective representation of chemical potentials \cite{bazant2013theory, bazant2017thermodynamic} and concentrations of inserted Li$^{+}$ in the electrode solid phase. The pulse waveforms and cell voltage recovery during the initial 10 s relaxation, or any cell voltage evolution in general, can be attributed to Li$^{+}$ chemical potential and concentration changes in the solid phase and electrolyte, which must involve both reaction kinetics and diffusion. As a result, rather than relying on unrealistically large pseudo-capacitances \cite{gomadam2008modeling}, it is more sensible to use mass transfer limitations at various stages of Li/CF$_x$-SVO cell operation to capture cell voltage transient behavior during high-rate pulses and post-pulse relaxations. While there is yet no consensus on the exact dependency of the diffusion 
coefficient of Li$^{+}$ in SVO on depth of discharge \cite{sauvage2010structural, takeuchi2001silver, ramasamy2006discharge, takeuchi1989lithium, lee2006electrochemical} from existing experimental studies of Li/SVO cells, these studies collectively showed that solid-phase diffusion in SVO is dependent on both operation conditions and SVO cathode filling fraction: (1) The internal resistance of Li/SVO was dominated by mass transfer limitations at higher current density pulses and longer pulse duration. (2) SVO as a cathode generally faces increasing diffusion limitations as more Ag$^{+}$ is displaced and V$^{5+}$ becomes only active material in the remaining oxide with high degree of disorder \cite{sauvage2010structural, crespi1995characterization, crespi2001modeling, west1995lithium}. However, note that in previous studies on solid-phase diffusion in SVO did not account for the fact that Li$^{+}$ can insert both into octahedral and tetrahedral sites of SVO lattice \cite{leifer2007nuclear} due to the parallel Ag$^{+}$ and V$^{5+}$ reduction reactions, respectively. As a result, we believe that using a single diffusion coefficient to represent mass transport of SVO overlooks the fact that Li$^{+}$ insertion takes place in parallel along two different pathways. Based on observations from existing experimental studies of SVO and similar modeling studies \cite{de2017explaining}, we hypothesize that solid-phase diffusion in SVO can be more accurately captured by two parallel diffusion processes across distinct pathways, separately for Li$^{+}$ insertion due to Ag$^{+}$ and V$^{5+}$ reactions.

SVO has a monoclinic unit cell \cite{zandbergen1994two, crespi1995characterization} under space group C$2\text{/}\text{m}$: vanadium oxide layers are formed by sharing edges and corners of distorted VO$_{6}$ octahedra, and the layered structure is stabilized by inserted Ag$^{+}$ between them \cite{leising1994solid}. In similar monoclinic SVO structures, the silver interlayer between the vanadium oxide sheets was found to allow for more facile lithium diffusion than the octahedral coordination of the vanadium \cite{cheng2011transition, liang2013synthesis, zhang2006synthesis}. The expectation that the Li$^{+}$ diffusion coefficient from the Ag$^{+}$ reaction $D_{_{\text{Li,Ag}}}$ is larger than Li$^{+}$ diffusion coefficient from V$^{5+}$ reaction $D_{_{\text{Li,V}}}$ matches the experimental measurements \cite{ramasamy2006discharge, strange2011physics} of solid-phase diffusion in SVO at different cell voltages, where the diffusion coefficient of Li$^{+}$ in SVO at $V_{\text{cell}}=3.2$ V (before Ag$^{+}$ depletion) was larger than that at $V_{\text{cell}}=2.6$ V (full Ag$^{+}$ depletion). In addition, for a Li/SVO cell under high-rate pulsing, considering the reaction of Ag$^{+}$ is much slower than V$^{5+}$, it is expected that it is mostly V$^{5+}$ reaction supporting the large current. As a result, it is also expected that the greater mass transfer limitations in Li/SVO cell measured at higher current density pulses and longer pulse duration can be mostly attributed to the lower $D_{_{\text{Li,V}}}$, which is supported by the measured or calculated poor Li$^{+}$ diffusivity in various vanadium oxides electrode materials \cite{xiong2008fabrication, wang2007synthesis, coustier1999doped, mcgraw1999li, lantelme2003electrochemical, machida1989behavior}. 

Based on the analysis above, Li$^{+}$ diffusion in SVO is described by two independent and parallel diffusion processes, 
\begin{equation}\label{eq:15}
\textbf{F}_{_{\text{Ag}}} = -\frac{D_{_{\text{Li,Ag}}}}{k_{\text{B}}T}c_{_{\text{Ag}}}\nabla \mu_{_{\text{Ag}}}
\end{equation}
\begin{equation}\label{eq:16}
\textbf{F}_{_{\text{V}}} = -D_{_{\text{Li,V}}}\nabla c_{_{\text{V}}}
\end{equation}
The flux of Li$^{+}$ due to Ag$^{+}$ reaction $\textbf{F}_{_{\text{Ag}}}$ is proportional to the gradient in chemical potential $\mu_{_{\text{Ag}}}$ because of its reaction's association with phase transition processes; Li$^{+}$ insertion due to V$^{5+}$ reduction is modeled as intercalation into solid solution material and thus the flux of Li$^{+}$ due to V$^{5+}$ reaction $\textbf{F}_{_{\text{V}}}$ is governed by Fick's law. For species conservation at different SVO particle depths, \begin{equation}\label{eq:17}
\frac{\partial c_{i}}{\partial t} = -\nabla \cdot \textbf{F}_{i}  \qquad \qquad i = \text{Ag},\text{V}
\end{equation}
\begin{figure}[!h]
  \centering
  \includegraphics[width = \columnwidth]{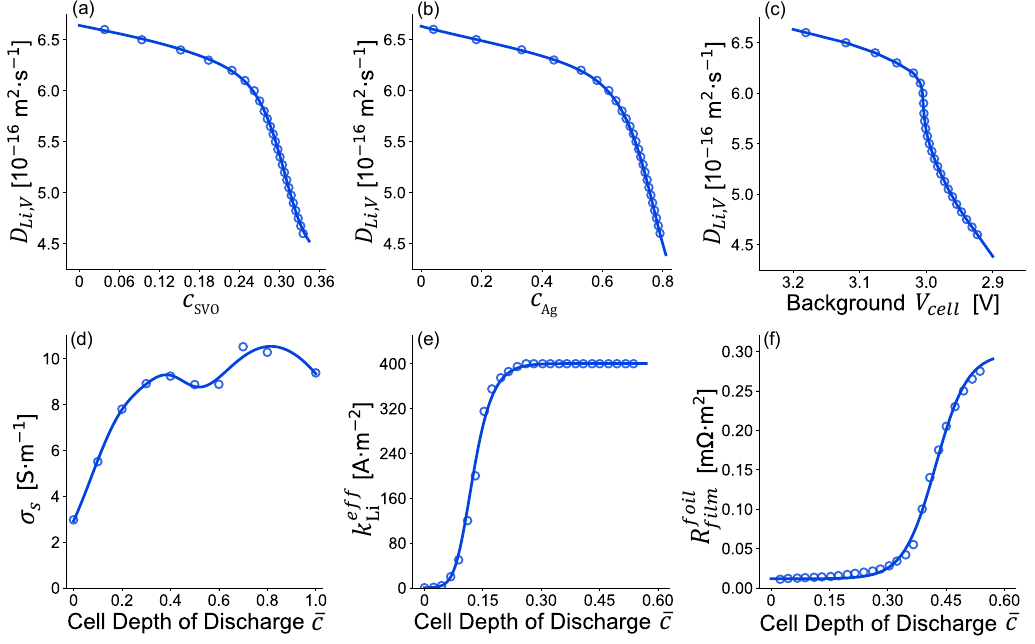}
  \vspace{-0.7cm}
  \caption{The empirical diffusion coefficient of Li$^{+}$ inserted due to V$^{5+}$ reduction reaction $D_{_{\text{Li,V}}}$ as functions of (a) SVO filling fraction $c_{_{\text{SVO}}}$,  (b) Ag$^{+}$ depth of discharge $c_{_{\text{Ag}}}$,  and (c) cell voltage at the last background monitoring mode timestep before each pulse. $D_{_{\text{Li,V}}}$ is in agreement with measured or calculated poor Li$^{+}$ diffusivity in various vanadium oxides electrode materials \cite{xiong2008fabrication, wang2007synthesis, coustier1999doped, mcgraw1999li, lantelme2003electrochemical, machida1989behavior}. 
  Additional empirical inputs to capture evolving electrode properties and degradation effects as functions of Li/CF$_x$-SVO cell depth of discharge $\overline{c}$ are (d) experimentally measured cathode electronic conductivity $\sigma_{s}$, (e) effective exchange current density rate constant $k^{\text{eff}}_{\text{Li}}$ that accounts for change in Li metal anode surface morphology after repeated pulsing, and (f) film resistance $R^{\text{foil}}_\text{film}$ to describe film resistance build-up at Li metal anode due to cathode dissolution. For $D_{_{\text{Li,V}}}$, $k^\text{eff}_{\text{Li}}$, and $R^\text{foil}_\text{film}$, each scatter point corresponds to fitted values that most accurately capture pulse waveform and post-pulse relaxation voltage at a specific pulses of the representative cell, with RMSE as a measure of goodness of fit.
  }
  \label{fig:combined_empirical}
\end{figure}
Under high-rate pulsing, due to the kinetic limitations of Ag$^{+}$ reaction, the impact of $D_{_{\text{Li,Ag}}}$ on cell voltage is overwhelmed by that of the smaller $D_{_{\text{Li,V}}}$; the impact of $D_{_{\text{Li,Ag}}}$ is also minimal after the pulse mode because the background monitoring mode operates under such small currents ($\sim$11\ $\upmu$A) that the system is not diffusion limited.
Our model is thus insensitive to $D_{_{\text{Li,Ag}}}$, and we use a generic solid-phase Li$^{+}$ diffusion coefficient $D_{_{\text{Li,Ag}}} = 5$$\times$$10^{-13} \  \text{m}^2$$\cdot$$\text{s}^{-1}$ \cite{ramasamy2006discharge}. For Li$^{+}$ intercalation from V$^{5+}$ reaction, as seen in Fig.\  \ref{fig:combined_empirical}(a), we use the pulse waveforms and initial 10 s of post-pulse relaxations of the representative cell to formulate an empirical $D_{_{\text{Li,V}}}(c_{_{\text{SVO}}})$ input to our model, with
\begin{equation}\label{eq:18}
\log_{10}( D_{_{\text{Li,V}}}(c_{_{\text{SVO}}})) = \frac{-15.424c_{_{\text{SVO}}}^5  -1.522c_{_{\text{SVO}}}^4 + 16.638c_{_{\text{SVO}}}^3  -12.637c_{_{\text{SVO}}}^2+ 3.982c_{_{\text{SVO}}} -0.466}{c_{_{\text{SVO}}}^5 + 0.114c_{_{\text{SVO}}}^4  -1.102c_{_{\text{SVO}}}^3 + 0.834c_{_{\text{SVO}}}^2 -0.263c_{_{\text{SVO}}}+ 0.031} 
\end{equation}
The $D_{_{\text{Li,V}}}$ values are also shown as functions of Ag$^{+}$ depth of discharge $c_{_{\text{Ag}}}$ and cell voltage in background monitoring mode before each pulse in Fig.\ \ref{fig:combined_empirical}(b)(c), respectively. We see that more drastic change of $D_{_{\text{Li,V}}}$ occurs when considerable fraction of Ag$^{+}$ has been depleted and cell voltage during background monitoring mode falls below $\sim$3.0 V. We hypothesize that the evolution of $D_{_{\text{Li,V}}}$ can be explained by the increased disorder, amorphization, and structural defects between vanadium oxide layers \cite{sauvage2010structural, takeuchi2001silver, onoda2001crystal, leising1994solid} caused by the size difference between Ag$^{+}$ and Li$^{+}$; the more Ag$^{+}$ displacement, the worse deterioration of the structural integrity of the crystalline SVO. While we assumed that the two Li$^{+}$ parallel but distinct insertion pathways coexist, since they share the same particle volume, it is possible that structural changes in SVO particles due to substantial Ag$^{+}$ displacement could affect the Li$^{+}$ intercalation pathways from V$^{5+}$ reaction, and subsequently impact $D_{_{\text{Li,V}}}$ evolution.

\subsection*{Evolution of CF$_x$-SVO cathode solid-phase conductivity}

It has been widely observed in the literature that formation of highly conductive metallic silver can considerably improve conductivity of SVO and silver vanadium phosphorus oxide (Ag$_2$VO$_2$PO$_4$) electrodes \cite{crespi1995characterization, crespi2001modeling, leising1994solid, ramasamy2006discharge, takeuchi2009electrochemical} and lower the cell resistance during operation. The increase in cathode conductivity, again, usually takes place during the first $\frac{1}{3}$ of cell depth of discharge, and is directly associated with the extent of Ag$^{+}$ depletion. 
While CF$_x$ has extreme low conductivity \cite{zhang2009electrochemical, nakajima1986discharge, nakajima1988discharge}, its reduction reaction can contribute to cathode conductivity change: the leftover carbon is a good conductive aid yet LiF is a insulator that forms inside both the carbon matrix and the free volume of the electrode \cite{read2011lif}. As a result, the evolution of the CF$_x$-SVO cathode conductivity is affected by the depth of discharge of active materials. An evolving cathode conductivity is thus another property that needs to captured, where smaller or larger cathode conductivity lowers and raises cell voltage during high-rate pulses, respectively, but both have limited impact on the pulse waveform shape.

For the specific CF$_x$-SVO hybrid cathode material in this study, its electronic conductivity is measured using the experimental configurations described in Gomadam et al.\  \cite{gomadam2003theoretical} at different cell depth of discharge as seen in Fig.\  \ref{fig:combined_empirical}(d). The evolution of the cathode conductivity $\sigma_{s}$ includes contributions from the aforementioned factors, and is passed into our model as a function of cell depth of discharge $\overline{c}$,
\begin{equation}\label{eq:19}
\sigma_{s}(\overline{c}) = 10.535\exp\!\left[{\left(\frac{\overline{c}-0.815}{0.535} \right)^{\!2}}\right] \!+ 4.878\exp\!\left[{\left(\frac{\overline{c}-0.205}{0.210} \right)^{\!2}}\right]\! + 1.462\exp\!\left[{\left(\frac{\overline{c}-0.396}{0.118} \right)^{\!2}}\right]
\end{equation}


\subsection*{Change in Li metal anode surface morphology}
Existing studies have observed  \cite{crespi2001modeling, crespi1995characterization, ramasamy2006discharge,root2011resistance} that Li metal anode's contribution to cell resistance could not be neglected in Li/CF$_x$-SVO and Li/SVO cells when tested over long periods of time. We noticed how previous assumptions \cite{liang2023hybrid, gomadam2007modeling, gomadam2009predicting} on Li metal anode having fast BV kinetics and thus negligible kinetic limitations might not apply to our pulse test. The quarterly high-rate defibrillation pulses have current densities reaching $\sim$$0.04 \ \text{A}$$\cdot$$\text{cm}^{-2}$), which is higher than the current densities ($\sim$$0.01 \ \text{A}$$\cdot$$\text{cm}^{-2}$) capable of causing significant Li stripping and pitting \cite{shi2018lithium} for the Li metal anode surface despite the presence of a solid electrolyte interphase (SEI) passivation layer. The surface damage from Li stripping and pitting was also found to be larger for higher current densities. As a result, the Li metal anode surface should not be always treated as an ideal rectangular cuboid or cylinder with even surfaces. The Li metal anode is damaged by repeated high current density pulses, ending with large amounts of $\sim \upmu\text{m}$ voids and stripped regions at its uneven surface, which yields a significant larger electrochemically active surface area \cite{shi2018lithium, boyle2021corrosion} compared to that of a pristine Li metal anode. 

The larger active area greatly boosts the intrinsic kinetics at Li metal anode and lowers cell resistance during high-rate pulsing, whose impact can again be captured by an area factor $f^\text{foil}_{A}$ and parameterized into the effective exchange current density rate constant $k^\text{eff}_{\text{Li}} = k_{\text{Li}}f^\text{foil}_{A}$ in the BV equation,
\begin{equation}\label{eq:20}
i_{_{\text{Li}}} = j_{_{\text{Li}}} e = k^\text{eff}_{\text{Li}} (c_{_{l}})^{0.5}\! 
\left[\exp\!\left(-\frac{0.5 e \eta_{_{\text{Li}}}}{k_{\text{B}}T}\right)\! - \exp\!\left(\frac{0.5 e \eta_{_{\text{Li}}}}{k_{\text{B}}T}\right) \right]
\end{equation}
where 
$k^\text{eff}_{\text{Li}}$ has been measured to be as high as $\sim$$400\ \text{A}$$\cdot$$\text{m}^{-2}$ in carbonate-based electrolytes during ultrafast cyclic voltammetry \cite{boyle2020transient}, which is substantially higher than the $k_{\text{Li}} \sim 5$$-$$30\  \text{A}$$\cdot$$\text{m}^{-2}$ measured for pristine Li metal anodes in similar electrolytes. Note that $k^\text{eff}_{\text{Li}}$ does not increase infinitely and tends to converge after enough cycles \cite{hobold2023beneficial}. In our case, we also expect $k^\text{eff}_{\text{Li}}$ to rapidly increase with the first few pulses and then reach a plateau because Li metal anode surface morphology change cannot be infinite. Under the same pulse current density, due to the increase in electrochemically active surface area, the average current density at the surface of Li metal anode decreases, and thus further repeating pulses at similar current density and pulse duration create lesser and lesser surface damage. As a result, there is a limit on how much anode surface morphology change takes place to accommodate for specific high-rate pulse current density and pulse duration. Since the quarterly defibrillation pulses always demand the very similar current density and all last for $\sim$4 s, as seen in Fig.\  \ref{fig:combined_empirical}(e), we formulate a sigmoid-type function to capture the impact of change in Li metal anode surface morphology on Li metal anode $k^\text{eff}_{\text{Li}}$,
\begin{equation}\label{eq:21}
\log_{10}k^\text{eff}_{\text{Li}}(\overline{c}) = \frac{2.6}{0.999+\exp(-8.670\overline{c}-25.943)} - \frac{1.2}{0.311+\exp(31.448\overline{c}-2.623)}
\end{equation}
where the $k^\text{eff}_{\text{Li}}$ plateau at $400 \ \text{A}$$\cdot$$\text{m}^{-2}$ is in reference to the measured effective exchange current density rate constants from Boyle et al.\  \cite{boyle2020transient}.

\subsection*{Cathode dissolution and deposition on Li metal anode}
Film resistance buildup \cite{crespi1995characterization, crespi2001modeling, leising1994solid, ramasamy2006discharge, root2011resistance} at the Li metal anode in Li/CF$_x$-SVO and Li/SVO cells has been widely observed in experimental studies \cite{bock2013kinetics, bock2013silver, bock2015structural, li2017synthesis,demayo2015cathode, bock2015mapping,li2017synthesis,le2022ag2v4o11}. The formation of multi-layered resistive films on the Li metal anode surface includes contribution from both traditional Li metal anode solid electrolyte interface (SEI) and deposition of dissolved vanadium and silver species from the cathode, whose presence was found via anode surface mapping using X-ray absorption spectroscopy and X-ray microfluorescence of the Li metal anode \cite{demayo2015cathode, bock2015mapping}. Cathode dissolution leads to loss of electroactive material to non-aqueous electrolyte \cite{demayo2015cathode, bock2015mapping, bock2013kinetics, bock2013silver, li2017synthesis}, where dissolved cathodic species are then able to travel to the anode and deposit. In most secondary batteries, cathode dissolution is considered as loss of active material \cite{kindermann2017sei, lin2013comprehensive, crawford2021lithium} and more emphasis is put on its impact on capacity fade. In our Li/CF$_x$-SVO batteries, since we have sufficient Li from the anode, we focus more on the impact of cathode dissolution on increasing cell resistance and diminishing pulse power capability of the cell. As a result, we introduce an additional film resistance term $R^\text{foil}_\text{film}$ to the effective overpotential $\eta^\text{eff}_{_{\text{Li}}} = \eta_{_{\text{Li}}} + i_{_{\text{Li}}}R^\text{foil}_\text{film}$ to replace $\eta_{_{\text{Li}}}$ in Eq.\ \ref{eq:22}. Assuming the buildup of film resistance at the anode has a limit, as seen in Fig.\  \ref{fig:combined_empirical}(f), we again formulate a sigmoid-type function to capture the growth of $R^\text{foil}_\text{film}$,
\begin{equation}\label{eq:22}
R^\text{foil}_\text{film}(\overline{c}) = \frac{2.867\!\times\!10^{-4}}{1+\exp\left(24.297\overline{c}-10.278\right)} + 2.983\!\times\!10^{-4}
\end{equation}

\section*{Results}

\subsection*{Model prediction of Li/CF$_x$-SVO battery pulse performance}

\begin{figure}[!h]
  \centering
  \includegraphics[width = \columnwidth]{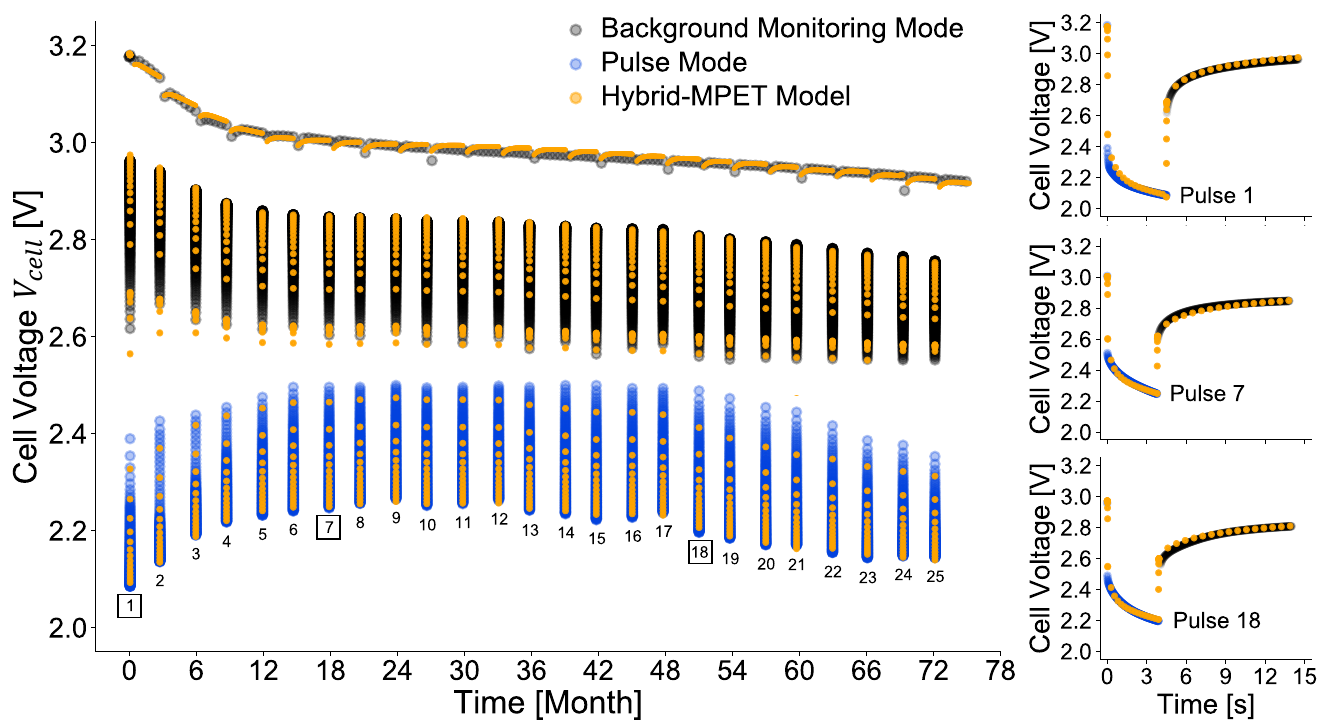}
  \vspace{-0.7cm}
  \caption{Comparison between experimental cell voltage measurements across pulse test data and predictions by its Hybrid-MPET model. The representative high-rate Li/CF$_x$-SVO cell is the one ranked 50$^{\text{th}}$ percentile by average cell pulse minimum voltage in the pulse test dataset. The black markers are measured cell voltage during background monitoring mode, and blue markers are the measured pulse waveforms of the battery. The orange markers show the Hybrid-MPET model predictions. The pulse waveform and initial 10 s of post-pulse relaxation from pulse 1, 7, and 18 are illustrated as examples, which are shown versus time that reference their respective pulse start times.}
  \label{fig:median_sample_results}
\end{figure}
As seen in Fig.\  \ref{fig:median_sample_results}, we overlay the pulse test experimental voltage measurements from the representative cell and its Hybrid-MPET model predictions; we also highlight the accuracy of our model in predicting battery pulse performance using the representative cell's pulse 1, 7, and 18. The corresponding cell voltage versus cell depth of discharge can be seen in Supplementary Fig.\ 3, and model predictions for all 25 pulses can be seen in Supplementary Fig.\ 4. The model is robust under vastly different time and current scales, and shows strong agreement with experimental cell voltage measurements across 3-month low-rate background monitoring mode segments, $\sim$4 s high-rate pulsing, and initial 10 s of post-pulse relaxation. Note at the start of a pulse, the experimental instant switching of $R_\text{ext}(t)$ from 270 k$\Omega$ to 0.65 $\Omega$ is normally difficult to replicate in a model; it involves many orders of magnitude transition of current and creates challenges for the Hybrid-MPET's differential algebraic equation (DAE) solver \cite{hindmarsh2005sundials}. At the end of the pulse, the same challenge exists for the switching of $R_\text{ext}(t)$ from 0.65 $\Omega$ back to 270 k$\Omega$. For numerical stability purposes, we introduce a stairway transition of $R_\text{ext}(t)$ that involves multiple 0.004 s timesteps, which are only $\sim$$0.1\%$ of most pulse durations. For visual clarity, some of these transition timesteps are removed from the cell voltage versus month plot in Fig.\ \ref{fig:median_sample_results}, but can be more clearly observed in the more zoomed in version of pulse waveform and initial 10 s relaxation plots in Fig.\  \ref{fig:median_sample_results} and Supplementary Fig.\ 4. Variable time stepping is also used to save computation time by keeping lower time resolution during 3-month background-monitoring and higher time resolution during pulsing and post-pulse relaxation.

\renewcommand{\thetable}{2}
\renewcommand{\arraystretch}{1.2}
\newcolumntype{P}[1]{>{\centering\arraybackslash}p{#1}}
\begin{table}[!h]
  \caption{RMSE as goodness-of-fit scores for multiple Li/CF$_x$-SVO battery pulse performance metrics}
  \centering
  \begin{tabular}{ccccccc}
  \hline
   Data & Pulse count $P$ & $V^{\text{min}}_{\text{pulse}}$ [V]& $V^{\text{avg}}_{\text{pulse}}$ [V]& $V^{\text{avg}}_{\text{relax}}$ [V]& $V^{\text{end}}_{\text{relax}}$ [V]
   \\
  \hline
    Representative cell &  25 & 0.007975 &  0.016980 & 0.010860 & 0.007877 
    \\
    Remaining 80 cells & 1496 & 0.031196 &  0.031694 & 0.015312 & 0.011557
    \\
    \hline
    \end{tabular}
  \label{table:RMSE}
\end{table}
Table \ref{table:RMSE} quantitatively compares our model predictions with experimental measurements using RMSE as a goodness-of-fit measure for describing Li/CF$_x$-SVO battery pulse performance. Our model, accurate in predicting pulse test performance of the representative cell, also generalizes well to the rest of the 80 cells in the dataset, obtaining $\text{RMSE}\approx 0.03$ V for both pulse minimum voltage $V^{\text{min}}_{\text{pulse}}$ and pulse average voltage $V^{\text{avg}}_{\text{pulse}}$. We also show residuals, i.e., the difference between experimental measurement and model prediction, of the four metrics in Supplementary Fig.\ 5; each of their distributions resembles a normal distribution with almost zero mean, which strongly infers that our model is representative of the high-rate Li/CF$_x$-SVO cells from the pulse test dataset and capable of predicting their pulse performance.

\subsection*{Active material contribution and post-pulse Li$^{+}$ redistribution}
\begin{figure}[!h]
  \centering
  \includegraphics[width = \columnwidth]{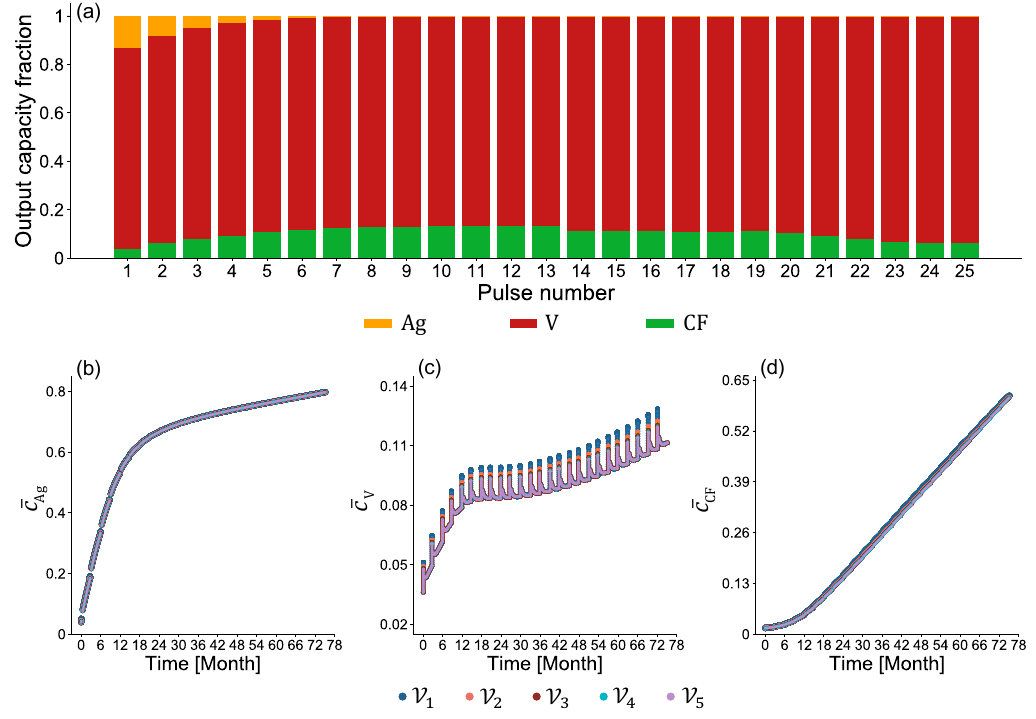}
  \vspace{-0.7cm}
  \caption{The model-predicted active material contributions to released capacity during high-rate pulses (a) and evolution of active material depth of discharge (b)(c)(d) for the representative cell, averaged across corresponding particle types within each finite volume $\mathcal{V}$. Finite volume $\mathcal{V}_{1}$ is the one closest to separator and finite volume $\mathcal{V}_{5}$ closest to the current collector.}
  \label{fig:contribution_evolution}
\end{figure}
\begin{figure}[!h]
  \centering
  \includegraphics[width = \columnwidth]{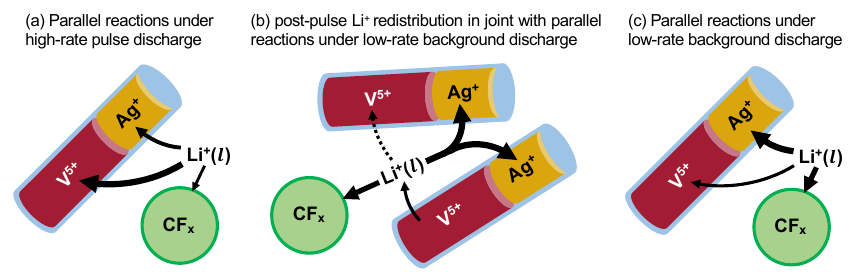}
  \vspace{-.7cm}
  \caption{Schematic drawing of reactions for each active material under different stages of cell operation: 
  (a) Under high-rate pulse discharge, parallel reactions co-exist and V$^{5+}$ reaction supports majority of high-rate cell current. 
  (b) Immediately after pulse, the cell is under low-rate background discharge. Due to the potential imbalances across active materials caused by change of active material compositions during pulsing, Li$^{+}$ redistribution takes place in joint with the parallel reactions. The process could be understood as CF$_x$ recharging SVO after pulsing from a macroscopic perspective. 
  (c) After the cell has been operating under low-rate background discharge for sufficiently long time, the active materials are at more balanced potentials and thermodynamics control reaction direction. Parallel reactions co-exist to support low-rate cell current, which is mostly supported by Ag$^{+}$ and CF$_x$ reaction. The sizes of arrows represent relative sizes of reaction rates, and the dashed arrow in (b) indicates that the intra-material and inter-particle Li$^{+}$ redistribution could happen if potential differences were sufficiently large.}
  \label{fig:Li_redistribution}
\end{figure}

For the multi-active material electrode CF$_x$-SVO, we are most interested in examining active materials contribution to pulsing current as well as their interactions. In Fig.\  \ref{fig:contribution_evolution}, to more concisely visualize these results, we show the model-predicted active materials contributions to released capacity during high-rate pulses and the evolution of active material depth of discharges throughout the test duration of the representative cell. $\overline{c}_{_{\text{Ag}}}$,$\overline{c}_{_{\text{V}}}$,$\overline{c}_{_{\text{CF}}}$ are obtained by averaging across corresponding particle types within each finite volume $\mathcal{V}$. Finite volume $\mathcal{V}_{1}$ is the one closest to separator and finite volume $\mathcal{V}_{5}$ closest to the current collector. We observe how V$^{5+}$ supports the majority of the pulse current across all pulses and the reduced contribution from Ag$^{5+}$ as it gradually becomes more depleted. Due to the constant resistive load operation mode, as soon as the pulsing mode is turned on, a decreasing cell voltage means that less current is required; the reaction rate and thus the reaction rate of each active materials changes at every timestep. To more clearly observe the change in reaction rates at the second timescale, we also show an expanded view of the model-predicted evolution of active material depth of discharge and their reactions rates during pulses 1, 7, 18 and corresponding initial 10 s of post-pulse relaxations in Fig.\ \ref{fig:neg_iv}. The reaction rates throughout the test duration of representative cell can be seen in Supplementary Fig.\ 6. Under high-rate pulsing, the outstanding spikes in V$^{5+}$ depth of discharge in Fig.\  \ref{fig:contribution_evolution}(c) and significant higher reaction rates in Fig.\  \ref{fig:neg_iv} again show the contribution from fast V$^{5+}$ reaction to support the majority of the pulse current. Despite their inherent slow reaction kinetics and limited contribution to supporting pulse current, Ag$^{+}$, CF$_x$ reactions are still active as seen in Fig.\ \ref{fig:contribution_evolution}(a) and Fig.\ \ref{fig:neg_iv}, and are driven by greater overpotentials after cell voltage drops significantly due to pulsing. The schematic picture describing the Ag$^{+}$, CF$_x$ reactions in parallel with a more dominant V$^{5+}$ reaction under high-rate pulsing can be seen in Fig.\  \ref{fig:Li_redistribution}(a). If the high-rate pulsing mode were kept for longer, V$^{5+}$ is expected to deplete first. Thus, our model predictions do not contradict the common experimental observations \cite{leising1994solid, crespi1995characterization, gomadam2007modeling, sauvage2010structural, ramasamy2006discharge, grisolia2011density} that, when Li/CF$_x$-SVO is under lower discharge rates, the active materials depletion order follows first Ag$^{+}$, then CF$_{\text{x}}$, and then V$^{5+}$. Rather, our modeling study offers more quantitative insights on how the utilization rate of different active materials is strongly dependent on operation conditions and the individual active material reaction kinetics.

During the initial 10 s of relaxation, as soon as the cell switches from pulsing mode to background monitoring mode, the cell voltage recovers and the demand of current decreases drastically. The overpotentials for the Ag$^{+}$, CF$_x$ reactions become very small, resulting in rapid decrease of $\overline{i}_{_{\text{Ag}}}$,$\overline{i}_{_{\text{CF}}}$. While three parallel reduction reactions across the active materials are still be taking place to support the low-rate background monitoring current, the spikes in V$^{5+}$ depth of discharge disappear after pulsing, and corresponds to the existence of $\overline{i}_{_{\text{V}}} < 0$ for finite volumes close to the separator in Fig.\  \ref{fig:neg_iv}. This indicates that some V$^{4+}$ is oxidized back to V$^{5+}$ at the cathode for some duration despite the overall cell being under discharge mode and net outward cell current. This phenomena corresponds to the Li$^{+}$ redistribution uniquely seen in multi-active material electrodes. We define Li$^{+}$ redistribution as: uniformly mixed active materials in a conductive solid phase are naturally in favor of thermodynamic balance, and will exchange Li$^{+}$ through the liquid electrolyte pathway to arrive at compositions at which the active materials have more balance potentials. Such phenomena has been observed experimentally from a more macroscopic perspective as CF$_x$ recharging the SVO \cite{chen2006hybrid} after high-rate pulses, and our model offers more comprehensive explanation. During a defibrillation pulse, rapid depletion of V$^{5+}$ close to the separator noticeably lowers the its potential and creates potential imbalance across the active materials; during initial 10 s of post-pulse relaxation and subsequent 3-month background discharge, the cell is held at such a low-rate current that it is nearly open circuit, and therefore thermodynamics control the reaction directions. 

\begin{figure}[!h]
  \centering
  \includegraphics[width = 14cm]{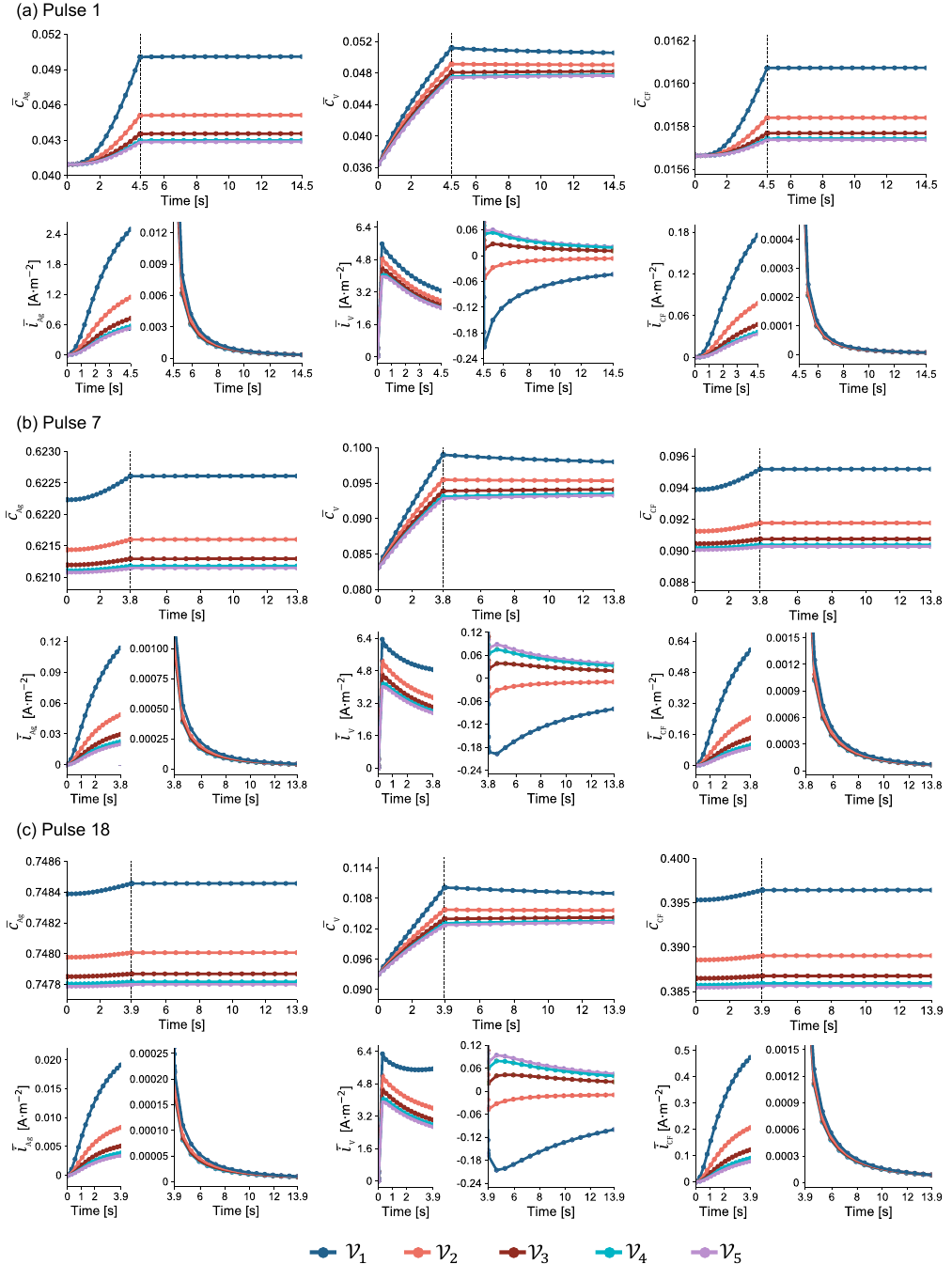}
  \vspace{-0.4cm}
  \caption{The model-predicted evolution of active material depth of discharge and their reaction rates for pulses 1, 7, 18 and their corresponding initial 10 s post-pulse relaxation. The depth of discharge and reaction rates are again averaged across corresponding particle types within each finite volume $\mathcal{V}$, where $\mathcal{V}_{1}$ is the one closest to separator and $\mathcal{V}_{5}$ closest to the current collector.}
  \label{fig:neg_iv}
\end{figure}

As seen in Fig.\ \ref{fig:Li_redistribution}(b), such process could be equivalently viewed as a previously inserted Li$^{+}$ due to V$^{5+}$ reaction leaves its site in SVO, diffuses to the SVO particle surface, enters the electrolyte, and then participates in either Ag$^{+}$, V$^{5+}$, or CF$_x$ reduction reaction that shift the compositions of active materials such that their potentials become more balanced. Note that the Li$^{+}$ redistribution can occur: (1) intra- or inter-particle, i.e., between active materials in the same particle or different particles; (2) intra- or inter-material, i.e., between the same or different active materials in the porous electrode, as long as there are potential imbalances. After a discharge pulse, if we consider relaxation under full open-circuit or near open-circuit low currents, the Li$^{+}$ redistribution process at the cathode involves multiple processes, including an electrochemical oxidation reaction for active materials at low potentials, Li$^{+}$ transport in solid, Li$^{+}$ transport in electrolyte, and a reduction reaction for active materials at higher potentials. As a result, during the initial 10 s of post-pulse relaxation as seen in Fig.\  \ref{fig:median_sample_results}, the rate of voltage recovery is collectively determined by pre-existing concentration gradients in solid and electrolyte, solid- and liquid-phase diffusion coefficients, and active material reaction rates at their respective compositions and overpotentials. Normally, the contribution from each physical process would be very difficult to decouple; yet in our model, because V$^{5+}$ reaction contributed most to the pulsing current and created Li$^{+}$ concentration gradient in SVO at the end of pulse, the the rate of post-pulse voltage recovery is found to be most sensitive to $D_{_{\text{Li,V}}}$, leading us to use $D_{_{\text{Li,V}}}$ to capture the pulse waveforms and initial 10 s of post-pulse relaxation.

Despite the introduction of quarterly pulses in the pulse test, their pulse duration is too short compared to the 3-month duration of near open-circuit low current background discharge. The previous reaction kinetics- and mass transport-limited system is soon overturned by thermodynamics control. The estimated diffusion time scales $\tau_{_{\text{Ag}}} = \frac{L^2}{D_{_{\text{Li,Ag}}}}$, $\tau_{_{\text{V}}} = \frac{L^2}{D_{_{\text{Li,V}}}}$ are both negligible compared to the timescale of months, leading to quickly diminishing Li$^{+}$ concentration gradients inside the SVO particles, which are shown in animated videos in Supplementary Information. The active materials reach more balanced potentials after considerable amount of Li$^{+}$ redistribution; the three reactions will take place in parallel as seen in Fig.\  \ref{fig:Li_redistribution}(c). The depth of discharge increase of V$^{5+}$ due to high-rate pulses are mostly reverted before the next pulse, keeping V$^{5+}$ at low compositions that not only allow more balanced potentials across the three active materials but also preserves the pulse capability of SVO material throughout battery operation. As a result, because low-rate background discharge between pulsing intervals is long enough, the general active material depletion order seen across Fig.\  \ref{fig:contribution_evolution}(b)(c)(d) matches those in medium-rate Li/CF$_x$-SVO cells \cite{leising1994solid, crespi1995characterization, gomadam2007modeling, sauvage2010structural, ramasamy2006discharge, grisolia2011density} and those directly predictable from active material OCV: first Ag$^{+}$, next CF$_{\text{x}}$, and then V$^{5+}$. 

\section*{Discussion}


We next discuss our newly learned insights on multi-active material electrode design principles after our modeling study. Conventionally, CF$_x$-SVO electrode has been viewed as a hybrid electrode that leveraged the great energy density of CF$_x$ to support years of low-rate background monitoring discharge and the power capability of SVO to support high-rate pulse currents. There has been very limited discussion on the interaction between active material interactions, and what functions these interactions would serve as under different cell current rates. Our model's demonstration of Li$^{+}$ redistribution adds a new layer of complexity in understanding multi-active material electrode design and operation: during the long periods of background monitoring, CF$_x$ is reduced to not only support the low-rate current after most Ag$^{+}$ depletion but also recharge SVO, whose $c_{_{\text{V}}}$ is at far-from-equilibrium composition due to V$^{5+}$ reaction's disproportionate contribution to the high-rate pulses. As a result, CF$_x$ has an additional function: it restores the power capability of SVO through the Li$^{+}$ redistribution mechanism, a process that is mostly thermodynamically driven. CF$_x$ not only directly contributes to the released capacity during pulsing as seen in Fig.\ \ref{fig:contribution_evolution}, but also indirectly contributes to maintaining the pulse performance of the cell by sustaining SVO. As a result, it should be expected that only a critical amount of SVO is needed to support pulse operations in Li/CF$_x$-SVO cells depending on pulsing duration, pulse frequency, and current density demanded. While the optimization of CF$_x$ and SVO capacity fractions is constrained by many more electrode and cell design requirements, we believe Li$^{+}$ redistribution will allow the cell retain most of its high-pulse pulsing capability with reduced the amount of SVO, which opens up additional space for tradeoff analysis and model-based optimization. 

It should be clear that, while Li$^{+}$ redistribution is inherently a thermodynamically driven process, it involves transport in solid, transport in liquid, as well as forward and back electrochemical reactions across different active materials. As seen in our study, while a continuous high-rate current pulse, at the cost of a large cell voltage drop, could continuously deplete V$^{5+}$, the existence and increase of potential imbalances across the active materials give rise to a stronger tendency or driving force for active materials to eliminate such potential differences. After a high-rate current pulse, if the cell is allowed to relax under open-circuit or near open-circuit low currents, Li$^{+}$ redistribution takes place and its rate is determined by either kinetics or transport depending on specific material property limitation. If the cell is kept under a high-rate current pulse for very long, the active materials would have compositions that yield large potential imbalances, effectively pushing to them to a far-from-equilibrium state. A previously inserted Li$^{+}$ due to V$^{5+}$ reaction would have increasing driving force to leave SVO (accompanied by V$^{4+}$ oxidizing back to V$^{5+}$), such that it could impede the V$^{5+}$ reduction required to support the required high-rate current. At this stage, the driving force for Li$^{+}$ redistribution across particles and active materials can act as an additional component of cell resistance during pulse mode. Based on the theoretical understanding above, besides existing practical considerations in electrode design, it is recommended to (1) optimize battery operation protocols such that the multi-active material electrode can avoid being at a far-from-equilibrium state before any high-rate current pulses, and (2) select active materials with similar thermodynamic properties, i.e. similar OCVs, when designing multi-active material electrodes for high-rate applications. 

We would also like to discuss our insights on parameterization of models for multi-active material electrode batteries. At the inter-particle scale, it has been common for models to capture distinct properties of active materials by parameterizing their thermodynamic, reaction kinetics, and mass transport properties separately. However, at the intra-particle scale, such as SVO containing both Ag$^{+}$ and V$^{5+}$ reactions and Li$^{+}$ occupying different lattice sites during insertion, such separate parameterization has been much rarer. Compared to the development of multi-active material electrode decades ago where the focus was on mixing particles of different types 
(inter-particle scale hybridness) to achieve better electrode performance, nowadays more research emphasis has been shifted towards creating single chemical compounds with multiple active materials 
(intra-particle scale hybridness) such as lithium manganese iron phosphate (LMFP) \cite{yang2021olivine, li2018structural, hu20163d, ombrini2023thermodynamics} with increased energy density and Si@C compounds \cite{hassan2014engineered, szczech2011nanostructured, feng2013facile, gao2009microporous, wang2014high} with more optimized structure design to naturally constrain volume expansion of Si. As a result, it is crucial for future electrochemical models to also separately parameterize the distinct properties of different active materials in the same particle to predict electrode performance under different operations. In the case of CF$_x$-SVO, the degree of distinctness between material properties combined with the quarterly high-rate current pulse test protocol provided enough separation to decouple individual property's impact on cell voltage. As a result, this work motivates the development of new electrochemical characterization techniques to separately probe material properties or designing unique operation conditions to decouple the properties' impact of cell voltage. Traditional methods to extract diffusion coefficient of Li$^{+}$ in solid materials like galvanostatic intermittent titration technique (GITT) would need additional updates since their property extraction is not only pulsing current rate dependent but also reliant on derivative of OCV with respect to state of charge, both of which become quite ambiguous in the presence of multiple active materials.

\vspace{-10pt}

\section*{Conclusion}
As a natural extension to our previous implementation of the Hybrid-MPET framework and modeling of medium-rate Li/CF$_x$-SVO and Li/SVO batteries, this work develops a physics-based model of Medtronic's high-rate Li/CF$_x$-SVO batteries that accurately predicts their performance under both low-rate background monitoring and high-rate defibrillation pulsing. We reflect distinct properties of active materials by parameterizing their thermodynamics, kinetics, and mass transport properties separately; use diffusion limitations of Li$^{+}$ in SVO to explain pulse and post-pulse relxation behavior of cell voltage; and introduce change in electrode conductivity, anode surface morphology, and film resistance to capture evolution of cell internal resistance over the course of the pulse test. Besides the rate dependence of active material utilization, our key insights are centered around the role of Li$^{+}$ redistribution in Li/CF$_x$-SVO and batteries with multi-active material electrodes: (1) Li$^{+}$ redistribution allows CF$_x$ to restore pulse capability of the CF$_x$-SVO electrode and indirectly contribute the released capacity during pulsing. (2) Li$^{+}$ redistribution can act as an additional component of cell resistance in batteries operating at far-from-equilibrium conditions. We also share our thoughts on how Li$^{+}$ redistribution affects formulation of operation protocols and design principles of batteries with other multi-active material electrodes, as well as how additional layers of complexity are introduced to porous electrode model parameterization and electrochemical experimental characterization due to parallel reaction and solid diffusion pathways. We hope that our models in the Hybrid-MPET framework can complement experimental research and accelerate the future development of hybrid electrodes with targeted performance.

Future extensions of this work could be focused on reducing the number of empirical inputs to model: introduce cathode conductivity change via percolation theories for multiple materials \cite{milton2022theory, stauffer2018introduction}, integrate cathode dissolution and anode film-resistance buildup into the model as irreversible side reactions \cite{chen2022porous, root2011resistance}. In conjunction with additional experimental measurement and characterization, it 
would be valuable to include effects from particle connectivity losses from increasing LiF presence, local temperature measurements, as well as electrode porosity and volume changes throughout battery testing. The generalizability of the model can also be further evaluated on battery pulse test where each quarterly pulse lasts longer to release more energy or where multiple pulses are performed consecutively every 3 months.

\vspace{-8pt}

\section*{Code and data availability}
Along with an example inputs to the model for the representative cell, the code for the Hybrid-MPET battery simulation software is available in the GitHub repository \cite{hybridMPETGithub}: 
\url{https://github.com/HarryQL/Hybrid-MPET/tree/medtronic_pulse}.
Additional example models for the other 80 cells can be provided upon request. Pulse test data of Li/CF$_x$-SVO batteries may be provided upon request.

\section*{Acknowledgements}
This work was supported by the Medtronic PLC.

\section*{Author Contributions}
\textbf{Qiaohao Liang}: Conceptualization, Methodology, Software, Validation, Formal analysis, Investigation, Data Curation, Writing - Original Draft, Writing - Review \& Editing, Visualization.
\textbf{Giacomo Galuppini}: Software, Data Curation, Writing - Review \& Editing. 
\textbf{Partha M. Gomadam}: Conceptualization, Resources, Data Curation, Writing - Review \& Editing. 
\textbf{Prabhakar A. Tamirisa}: Conceptualization, Resources, Data Curation, Writing - Review \& Editing, Supervision, Project administration.
\textbf{Jeffrey A. Lemmerman}: Resources, Data Curation.
\textbf{Michael J. M. Mazack}: Writing - Review \& Editing.
\textbf{Melani G. Sullivan}: Resources.
\textbf{Richard D. Braatz}: Writing - Review \& Editing, Supervision, Project administration.
\textbf{Martin Z. Bazant}: Conceptualization, Methodology, Writing - Review \& Editing, Supervision, Project administration.

\section*{Supplementary Information}
Supplementary Information is available with the electronic version of the article on the journal's website. 

\section*{Competing interests}
The authors declare that they have no known competing financial interests or personal relationships that could have appeared to influence the work reported in this paper.


\printbibliography

\vspace{+20pt}

\end{document}